\documentclass{aa}  

\usepackage[colorlinks=true, linkcolor = blue,
            urlcolor  = blue,
            citecolor = blue,
            anchorcolor = blue, bookmarks=false]{hyperref}
\usepackage{graphicx}
\usepackage{txfonts}
\usepackage{multirow}
\usepackage{natbib}
\usepackage{enumitem} 
\usepackage{color}
\usepackage{threeparttable}
\usepackage{tikz,amsmath}
\usetikzlibrary{shapes.geometric, arrows}
\usepackage{ulem}
\usepackage{wrapfig}
\usepackage{rotating}
\usepackage{lscape} 
\tikzstyle{process} = [rectangle, minimum width=1.5em, minimum height=3.5em, text centered, draw=blue, fill=gray!10]
\tikzstyle{process2} = [rectangle, minimum width=1.5em, minimum height=3.5em, text centered, draw=white, fill=white]
\tikzstyle{arrow} = [thick,->,>=stealth]
\usepackage{cancel}
\usepackage{ulem}
\bibpunct{(}{)}{;}{f}{}{,} 

\usepackage[colorlinks=true,citecolor=blue,urlcolor=blue]{hyperref}

\newcommand{\teff}{$T_{\mathrm{eff}}$}

\newcommand{\logg}{\mbox{log \textit{g}}}

\newcommand{\titan}{\textsc{Titans}}

\begin{document}

   \title{The nucleosynthesis of Ba in the Early Universe
   \thanks{Based on observations made with ESO Telescopes at the La Silla Paranal Observatory.}
   }

   \subtitle{Constraints from elemental abundances and isotopic ratios}

   \author{R. E. Giribaldi\inst{1}\thanks{Corresponding author: riano.escategiribaldi@inaf.it;\\ rianoesc@gmail.com }
          \and
          M. Palla\inst{1}
           \and
          L. Magrini\inst{1}
          \and
          F. Rizzuti\inst{3}
          \and
          G. Cescutti\inst{3}
          \and
          D. Vescovi\inst{2,4}
          \and     
          S. Cristallo\inst{2,4}
          \and
          M. T. Belmonte\inst{4}
          \and
          S. Randich\inst{1}
          }
   \institute{INAF – Osservatorio Astrofisico di Arcetri, Largo E. Fermi 5, 50125
Firenze, Italy \and 
              INAF - Osservatorio Astronomico d\textquotesingle Abruzzo, Via Maggini snc, Teramo, Italy 
              \and
              Dipartimento di Fisica, Sezione di Astronomia, Universit\`{a} di
Trieste, Via G. B. Tiepolo 11, 34143 Trieste, Italy 
\and
              INFN - Sezione di Perugia, Via A. Pascoli, 06123 Perugia, Italy
\and
              Universidad de Valladolid, Departamento de Física Te\'{o}rica, At\'{o}mica y Optica, P.
de Bel\'{e}n, 7, Valladolid 47011, Spain              
             }
             
    \date{Received NN, 2026; Accepted NN, 2026}

 
  \abstract
{The [Ba/Eu] abundance ratio is commonly adopted as a tracer of the relative contributions of the slow (s) and rapid (r) neutron-capture processes. However, at low metallicity ([Fe/H] $\lesssim -2$ dex), barium can be produced efficiently by both processes, rendering [Ba/Eu] non-deterministic. We propose to use barium isotopic ratio, from the fitting of resonance \ion{Ba}{ii} line profiles affected by hyperfine splitting. This approach requires precise atomic and stellar parameters,  together with advanced spectral modelling, which, so far, remained insufficiently validated.}
{To provide a robust prescription of line-profile modelling for a reliable determination of the s- and r-processes fractions of barium in  ordinary and peculiar stars. These observational determinations can be used to place constrains on Galactic chemical evolution models, testing yields from different astrophysical sources.}
{We assess the performance of 1D~LTE and 1D~non-LTE synthesis, and 3D~non-LTE abundance corrections to model Ba lines.
Alongside barium abundances and its isotopic ratios, we determine Eu and other neutron-capture element abundances to validate the method in the \titan\ metal-poor benchmark stars.
The  observational results are compared with the predictions of stochastic Galactic chemical evolution models that account the inhomogeneous mixing in the early  times.}
{We find that 1D LTE and 3D non-LTE Ba abundance determinations are equivalent, whereas
the 1D non-LTE approach leads to systematic underestimations.
{These underestimations bias isotopic fractions toward higher r-process contributions.}
The inferred s- and r-process fractions demonstrate that [Ba/Eu] alone is an ambiguous tracer for ordinary stars within the range $-0.8 \lesssim$ [Ba/Eu] $\lesssim 0$~dex. However, we identify an observational pattern in the [Ba/Eu]-[Ba/Fe] plane that can distinguish, at a given [Ba/Fe], the dominant neutron capture process. 
The comparison of our set of models, both for the proto-Milky Way halo and for Gaia-Enceladus galaxy is used to put constraints on the production of Ba at low metallicity, especially evaluating the role of  rotating massive stars. }
{ The method here developed can be applied with confidence to both ordinary stars and peculiar stars enhanced in barium.}


   \maketitle
%

\section{Introduction}

High-resolution spectroscopy provides a unique opportunity to probe the products of stellar nucleosynthesis not only through elemental abundances, but also through isotopic ratios \citep[e.g.][]{Clayton1968psen.book.....C, Sneden2008ARA&A..46..241S} Unlike atomic abundances, which  trace the cumulative outcome of multiple nucleosynthetic channels, isotopic fractions retain a more direct imprint of the physical conditions under which elements were formed. Establishing the connection between isotopic diagnostics and elemental abundance ratios is therefore essential to assess whether commonly used abundance ratio indicators are reliable proxies of the underlying nucleosynthetic processes \citep[see, e.g.][]{Mashonkina2010A&A...516A..46M}.

Among the heavy elements, those produced by neutron-capture reactions are particularly powerful tracers of nucleosynthesis processes and Galactic chemical evolution. The rapid neutron-capture process (r-process), responsible for the synthesis of roughly half of the elements heavier than iron, remains one of the major open issue in nuclear and stellar astrophysics \citep{Thielemann2017ARNPS..67..253T}. Several astrophysical sites have been proposed, including core-collapse supernovae, neutron-star winds, and neutron-star mergers
\citep[e.g.][]{argast2004A&A...416..997A,wanajo2006NuPhA.777..676W,panov2009A&A...494..829P,Thielemann2017ARNPS..67..253T}. While early theoretical work favoured massive-star explosions as r-process sites \citep[see][]{Van2023A&A...670A.129V}, the detection of heavy-element absorption features in the optical counterpart of a neutron-star merger \citep{Watson19} has renewed interest in compact binary mergers as major contributors to r-process nucleosynthesis.

Metal-poor stars offer a direct window into these early enrichment processes. At low metallicity, the chemical inventory of stars is expected to be dominated by a limited number of nucleosynthetic events \citep[e.g.][]{Hartwig2019MNRAS.482.1204H}, and r-process signatures are particularly prominent. 
Europium, being almost purely produced by the r-process \citep[e.g.][]{Bisterzo14,Prantzos20}, provides a robust reference element, whereas barium presents a more complex behaviour: although it is primarily produced by the s-process at solar metallicity, the contribution from the r-process increases substantially at lower metallicities \citep[e.g.][]{mashonkina2006A&A...456..313M}.

In this context, barium isotopic fractions  (or  isotopic ratios) offer a unique diagnostic. 
Unlike its elemental abundance, which primarily trace the integrated r- and s-process contribution, Ba isotopes encode information on the neutron-density regime and therefore on their putative sources. On the one hand, even isotopes ($^{134}$Ba, $^{136}$Ba, $^{138}$Ba) are primarily produced by s-process events in low-mass Asymptotic Giant Branch (AGBs) and massive stars, with $^{134}$Ba and $^{136}$Ba being s-only isotopes. 
Conversely, the fraction of odd Ba isotopes ($^{135}$Ba, $^{137}$Ba) encodes information on the r-process contribution, as mostly produced in high-neutron density events.
The Ba odd-isotope fraction, first measured in metal-poor stars by \cite{magain1993A&A...268L..27M}, has therefore been  used to quantify the relative contributions of the r- and s-processes and to constrain the nature of early r-process sources \citep[e.g.][]{meng2016A&A...593A..62M, cescutti2021A&A...654A.164C, Sitnova25, sitnova2025A&A...704A.103S, giribaldi2025A&A...702A..65G,giribaldi2026arXiv260511074G}.

From both observational and spectroscopic perspectives, barium remains a particularly challenging element. Its strong resonance and subordinate lines are highly sensitive to microturbulence and are  believed to exhibit non-local thermodynamic equilibrium (NLTE) effects \citep{sitnova2025A&A...704A.103S}. In addition, its isotopic composition leaves only subtle signatures in line profiles. 
In this work, we aim to demonstrate that the inferred Ba isotopic ratios strongly depend on the adopted modelling assumptions, thereby highlighting the importance of carefully assessing the analysis methodology.
We revisit the problem of barium abundances and isotopic fractions using high-resolution, high signal-to-noise spectra of metal-poor ([Fe/H]\footnote{[A/B] = $ \log{\left( \frac{N(\text{A})}{N(\text{B})} \right )_\text{Star}} - \log{\left( \frac{N(\text{A})}{N(\text{B})} \right )_\text{Sun}} $, where $N$ denotes the number abundance of a given element.} $\lesssim -1$~dex) dwarf and giant stars. We combine the analysis of Ba with that of europium, adopted as a reference r-process element.
Our goal is twofold: first, to identify a robust strategy for deriving  accurate Ba abundances and isotopic fractions; second, to assess whether elemental abundance ratios provide reliable proxies for the relative contributions of the s- and r-processes in stellar material.
 The method we present is required for peculiar stars enhanced in barium ([Ba/Fe] $\gtrsim 0.70$~dex), where the contributions of the nucleosynthetic processes on their chemical patterns is difficult to determine solely by the abundance themselves.
For this reason we first validate it with a relatively large sample of ordinary stars.
Finally, we compare our results with a set of stochastic chemical evolution models to place the derived abundances and isotopic ratios on a theoretical footing, and to gain further insight into the different sources of neutron-capture enrichment in the early Universe.\\

The paper is organised as follows. In Sect.~\ref{sec:data}, we present the observational data  and atmospheric parameters of our stellar sample. In Sect.~\ref{sec:abuns} we describe our derivations of Eu, Ce, La, and Sr.
In Sect.~\ref{sec:barium_modelling} we describe the methodology for modelling barium lines, and provide the atomic data adopted to derive abundances and  isotopic fractions. In the same section, we assess the performance of 1D~LTE and 1D~NLTE approaches. 
The cases of peculiar stars are discussed in detail. 
In addition, we provide ranges of data quality and stellar parameters for an efficient application of the method. 
In Sect.~\ref{sec:discusion}, we employ our Galactic chemical evolution models to interpret the observed chemical signatures. Finally, in Sect.~\ref{sec:conclussions}, we summarise our results and present our conclusions.

\section{Data, sample, and atmospheric parameters}
\label{sec:data}

Our sample of  68 ordinary stars is listed in Table~\ref{tab:barium} and consists of F- and G-type metal-poor stars. The first subset comprises dwarf stars analysed by \citet{giribaldi2021A&A...650A.194G}, referred to as the \titan~I sample. All stars in this subset have high signal-to-noise (S/N) co-added spectra  of resolution $R = \lambda/\Delta \lambda > 40\,000$ obtained with the Ultraviolet and Visual Echelle Spectrograph \citep[UVES;][]{2000SPIE.4008..534D}, High Accuracy Radial velocity Planet Searcher \citep[HARPS;][]{mayor2003Msngr.114...20M}, or Echelle Spectrograph for Rocky Exoplanets Search and Stable Spectroscopic Observations \citep[ESPRESSO;][]{Pepe2021A&A645A96} instruments; see data in Table~\ref{tab:parameters}.
The metal-poorest tail ([Fe/H] $\lesssim -2$~dex) is probed using red giants from \citet{giribaldi2023A&A...679A.110G}, referred to as the \titan~II sample, as barium lines of dwarfs in this metallicity range become undetectable. These stars  also have spectra of $R > 40\,000$ and  are listed in the second part of Table~\ref{tab:barium}. We selected only those with  public spectra in the European Southern Observatory (ESO) phase~3 archive\footnote{\url{https://archive.eso.org/wdb/wdb/adp/phase3_main/form}} that exhibit detectable subordinate (6141 and 6496~\AA) and resonance (4934~\AA) \ion{Ba}{ii} lines.

The Galactic membership of the stars is listed in Table~\ref{tab:parameters}. For the \titan~I stars, it was established in \cite{giribaldi2023A&A...673A..18G}. The \textit{in situ} populations considered in this study include the Milky Way discs, the Splash population \citep[][also referred to as the heated disc]{belokurov2020MNRAS.494.3880B}, and Erebus population, which was previously identified as Thamnos 2 and interpreted as an \textit{ex situ} structure by \cite{Koppelman2019A&A...631L...9K}, but here is treated as \textit{in situ} based on stellar age evidence from \cite{giribaldi2023A&A...673A..18G}. The only \textit{ex situ} population is the Gaia–Enceladus–Sausage merger \citep[GES;][]{brook2003ApJ...585L.125B,Belokurov2018,helmi2018}.
The giant stars of \titan~II are all associated with the metal-poor tail of the Milky Way and are interpreted as having an \textit{in situ} origin \citep[proto-MW henceforth, ][]{giribaldi2025A&A...698A..11G}. These classifications are adopted in Sect.~\ref{sec:discusion} to ensure a fair comparison with Galactic chemical evolution models.

To validate the method with peculiar stars, we include two well studied r-process dominated benchmarks with UVES spectra: 
BPS~CS~22892-052 \citep[hereafter Sneden star,][]{Sneden94} and BPS~CS~31082-001
\citep[hereafter Hill star, ][]{hill2002A&A...387..560H}.
In addition, we include
two carbon-enhanced metal-poor (CEMP\footnote{According to the classification of \citet{beers2005ARA&A..43..531B}, which requires [C/Fe]~$> 1$~dex.}) stars with both r- and s-process signatures according to the literature.
One is TYC~6044-714-1, with a UVES spectrum, which is a s+r processes star \citep{gull2018ApJ...862..174G,giribaldi2026arXiv260511074G}. Another is HD~196944 with a HARPS spectrum, for which the influence of the s-, r-, and intermediate \citep[i-][]{cowan1977ApJ...212..149C} processes remain under debate \citep{abate2015A&A...581A..22A,sitnova2025A&A...704A.103S}.
The CEMP category also includes the Sneden star.
All these categories are indicated in Table~\ref{tab:parameters}.
The cases of these stars are analysed in detail in the Appendix Sects.~\ref{sec:hd196944} and \ref{sec:peculiars}. The case of TYC~6044-714-1 is presented in detail in \cite{giribaldi2026arXiv260511074G}. 
In addition, for validation purposes, we include a solar spectrum reflected by Ganymede ($R = 86\,000$) retrieved from the MELCHIORS public database\footnote{\url{https://www.royer.se/melchiors.html}.}
\citep{Royer2024A&A...681A.107R}.

The atmospheric parameters adopted for our sample are listed in Table~\ref{tab:parameters}. Effective temperatures (\teff) for the majority of the stars were refined by \citet{giribaldi2025A&A...698A..11G}, and we adopt those values here. For stars not included in that analysis, \teff\ is taken as the mean of the determinations from H$\alpha$ fitting and either the infrared flux method \citep[IRFM;][]{Casagrande2010} or the photometric calibrations based on the IRFM presented in Tables~1 and 2 of \citet{giribaldi2021A&A...650A.194G} and \citet{giribaldi2023A&A...679A.110G}, respectively.
Although Table~\ref{tab:parameters} reports internal uncertainties, we note that the total uncertainty in \teff—including model accuracy—is estimated to be 40~K for both dwarfs and giants.

For dwarf stars, surface gravities (\logg) and [Fe/H] are adopted from \citet{giribaldi2021A&A...650A.194G}. 
Their microturbulent velocity ($v_{mic}$) was determined by the excitation equilibrium of Fe lines from fixed isochronal \logg; these unpublished quantities are provided in Table~\ref{tab:parameters}.
For giants, [Fe/H] and $v_{mic}$ are determined from the excitation equilibrium of Fe lines under 1D~NLTE, while surface gravities for giants are taken as the average of the values derived by \citet{giribaldi2023A&A...679A.110G}, based on fitting the Mg I triplet, and those obtained from the excitation and ionization equilibrium of Fe lines. The associated uncertainty is estimated as the standard deviation of the differences between these two \logg\ determinations. The atmospheric parameters of TYC~6044-714-1 were determined using a similar methodology \citep{giribaldi2026arXiv260511074G}.
Line synthesis is performed with the radiative transfer code Turbospectrum \citep{gerber2023} and MARCS model atmospheres \citep{gustafson2008}. For 1D~NLTE calculations, we adopt departure coefficients based on the model atoms developed by \citet{bergemann2012MNRAS.427...27B} and \citet{semenova2020A&A...643A.164S}. We use the same Fe line list compiled by \citet[][hereafter Paper~I]{giribaldi2025A&A...702A..65G} , based on lines from \citet{jofre2014A&A...564A.133J} and \citet{melendez2009A&A...497..611M}. Atomic data are adopted from \citet{heiter2021A&A...645A.106H}.

\begin{figure}
    \centering
    \includegraphics[width=0.85\linewidth]{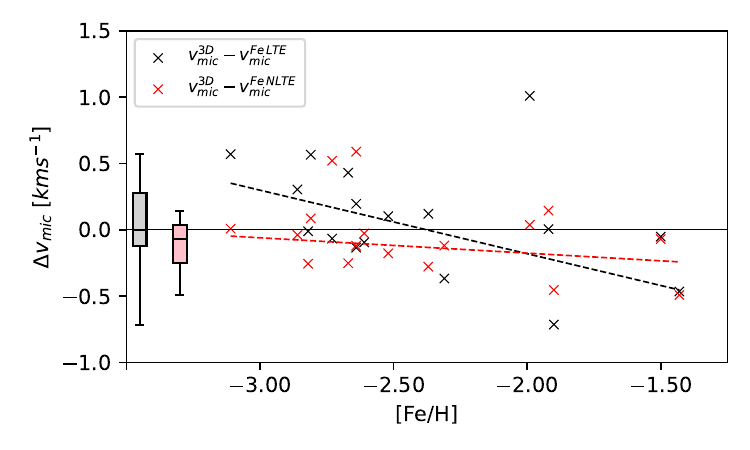}
    \caption{\tiny Difference between 3D~LTE model-based microturbulence and $v_{mic}$ from Fe lines under 1D~LTE (black crosses) and 1D~NLTE (red crosses).
    Dashed lines of corresponding colours represent the trends of the dispersions.
    Boxes of corresponding colours represent 25, 50 (median), and 75\% quantiles.
    }
    \label{fig:vmic_feh}
\end{figure}

Microturbulence plays a critical role in our determination of  the isotopic ratios inferred from Ba resonance lines in giant stars, both directly and indirectly through its effect on the derived barium abundance. To minimise associated uncertainties, we adopt the strategy of Paper~I, which showed that microturbulent velocities derived from Fe lines are in good agreement with the 3D~LTE model–based calibrations of \citet{2016A&A...585A..75D}.
In Fig.~\ref{fig:vmic_feh} we compare 1D~LTE and 1D~NLTE outcomes ($v_{mic}^{\mathrm{Fe\,LTE}}$ and $v_{mic}^{\mathrm{Fe\,NLTE}}$) with the 3D~LTE-based estimates. 
The 1D~NLTE results show clearly a reduced scatter, particularly in the most metal-poor regime where the number of usable lines is limited. The pink box in the figure indicates the 25th and 75th percentiles of the differences between the 3D~LTE-calibrated estimates and $v_{mic}^{\mathrm{Fe\,NLTE}}$ ($-0.25$ and $+0.04$ km~s$^{-1}$, respectively), with a median offset of $-0.07$ km~s$^{-1}$. We therefore adopt microturbulence values obtained by averaging them. The associated uncertainty is taken as the half-range between the 25th and 75th percentiles relative to the median, corresponding to $\sigma(v_{mic}) = \pm_{0.2}^{0.1}$ km s$^{-1}$.

\section{Abundances of Eu, Ce, La, and Sr}
\label{sec:abuns}

As a  propaedeutic step, we measured carbon abundances by fitting the CH G band near 4300~\AA, when blue spectral coverage is available. For CEMP giants, this region is avoided due to saturation effects, and the Swan $^{12}\mathrm{C}_2$ band at 5140–5166~\AA\ is instead employed.
Because $\alpha$-elements can affect carbon abundance determinations, [Mg/Fe] ratios from \cite{giribaldi2025A&A...698A..11G} are included in the spectral synthesis. When unavailable, magnesium abundances are derived from the Mg I lines at 5528 and 5711~\AA\ under 1D~NLTE, following the same prescriptions.

Europium abundances\footnote{
We use the notation A(X), which, for a given element X, the logarithmic absolute abundance is
defined as the number of atoms of element X per $10^{12}$ hydrogen
atoms, log $\epsilon$(X) = A(X) $\equiv$ log$_{10} (NX/NH) + 12.0$.} are derived from the 4129.72 and 4205.03~\AA\ lines under 1D~NLTE, using the coefficients of \citet{storm2024}; these are resported in Table~\ref{tab:barium}. For stars with two usable lines, the internal uncertainty is taken as the standard deviation of the derived values (typically 0.05~dex for dwarfs and 0.03~dex for giants); for stars with only one line, we adopt the line-fitting uncertainty. This error is added in quadrature to the abundance variations induced by uncertainties in \teff, \logg, $v_{mic}$, and [Fe/H] to obtain the total uncertainty. To estimate these effects, each atmospheric parameter was varied individually while keeping the others fixed. 
For both dwarfs and giants, uncertainties in $v_{mic}$ and [Fe/H] are negligible. The effect of \logg\ is negligible for dwarfs, whereas for giants it affects the abundances by $\sim$0.05~dex per 0.15~dex in \logg. The sensitivity to \teff\ is similar in both dwarfs and giants, with abundances changing by less than $\sim$0.04~dex per 40~K in \teff.

Ce, La, and Sr abundances were determined under 1D ~LTE from the lines listed in Table~\ref{tab:atomic_parameters} and are reported in Table~\ref{tab:barium}. Atomic parameters were adopted from \cite{heiter2021A&A...645A.106H} for $\lambda \geq 4200$~\AA\ and from the VALD3 database \citep{Ryabchikova2015PhyS...90e4005R} for $\lambda < 4200$~\AA. The uncertainty treatment follows that described for Eu. The sensitivities of these abundances to uncertainties in the atmospheric parameters are nearly equivalent to those found for Eu \citep[e.g. Fig.~D.3. in][]{giribaldi2026arXiv260511074G}.

Eu, Ce, La, and Sr abundances were derived to examine the behaviour of Ba relative to other neutron-capture elements.
The [Ba/Eu] ratios are discussed in Sect.~\ref{sec:fiducial} to assess which Ba line-formation approach, 1D~LTE or 1D~NLTE, provides more reliable abundances. The [Ba/Ce], [Ba/La], and [Ba/Sr] ratios are analysed in Sect.~\ref{sec:other_el}.

\section{Barium Line Modelling}
\label{sec:barium_modelling}

The determination of barium abundances is strongly affected by the choice between 1D~LTE and 1D~NLTE modelling. In the following sections, we demonstrate that the 1D~NLTE approach systematically underestimates $A(\mathrm{Ba})$ by up to $\sim$0.4~dex, depending on the abundance or line strength. 
Our tests conducted on the
solar spectrum and our stellar sample indicate that the 1D LTE approach generally yields values close to the true abundances.
This is, with a zero point tied to the chondritic solar value A(Ba) $= 2.18 \pm 0.02$~dex \citep{lodders2009LanB...4B..712L}.
Conversely, 1D~NLTE modelling reproduces the observational line at 4934~\AA\ with higher fidelity than 1D~LTE in metal-poor giants.
We therefore adopt a hybrid approach: 1D~LTE-derived A(Ba) values are used to fix the abundance, while 1D~NLTE calculations are employed to model the 4934~\AA\ line for the derivation of isotopic ratios. This strategy enables us to obtain reliable isotopic diagnostics.
Details on isotopic fraction biases from adopting either pure 1D~LTE or pure 1D~NLTE modelling are given in Sect.~\ref{sec:NLTE}.

\subsection{Isotopic fractions of barium: AGB vs. massive rotating stars}
\label{sec:iso_theoric}

A key ingredient for the determination of barium isotopic ratios is the assumed isotopic mixture, which we adopt from theoretical nucleosynthesis predictions. The relative contributions of the various barium sources change across different epochs, leading to different isotopic mixtures. Properly accounting for these variations is crucial to disentangle the nucleosynthetic origin of barium in the Galaxy, including the main s-process in AGB stars and the weak s-process in massive stars. For the latter, rotation-induced mixing plays a pivotal role, especially at low metallicities. In this section, we aim to identify the isotopic mixture that provides the most appropriate input for our determination of the barium isotopic ratios.

Table \ref{tab:ratios} summarizes the relative isotopic fractions of Barium. 
{These quantities indicate the normalised fraction to which each Ba isotope is produced.}
Theoretical yields for non-rotating AGB stars at metallicities $\rm [Fe/H] = -2.14, -1.14$, and $-0.14$ are retrieved from the MARTINI platform\footnote{\url{https://martini.oa-abruzzo.inaf.it/}}~\citep{Cristallo15,Bezmalinovich26}. 
Yields for massive stars at $\rm [Fe/H] = 0, -1, -2$, and $-3$ are taken from the ORFEO database\footnote{\url{https://orfeo.oa-roma.inaf.it/}}~\citep{Limongi18}. For the massive star models, we consider three initial rotation velocities: non-rotating ($v_{rot} = 0$~km~s$^{-1}$), intermediate ($v_{rot} = 150$~km~s$^{-1}$), and fast rotators ($v_{rot} = 300$~km~s$^{-1}$). The pure s- and r-process isotopic fractions of the Solar System are also provided for direct comparison.
\begin{table}[!htpb]
\caption{Barium isotopic fractions}
\label{tab:ratios}
\centering
\tiny 
\begin{threeparttable}
\begin{tabular}{lc|ccccc}
\hline\hline
Source & [Fe/H] & $^{134}$Ba & $^{135}$Ba & $^{136}$Ba & $^{137}$Ba & $^{138}$Ba \\
\hline
AGB & $-0.14$ & 0.0215 & 0.0236 & 0.0689 & 0.1508 & 0.7353 \\
 ($v_{rot}=0$) & $-1.14$ & 0.0169 & 0.0175 & 0.0573 & 0.0770 & 0.8313 \\
& $-2.14$ & 0.0352 & 0.0221 & 0.0999 & 0.0853 & 0.7574 \\
\hline
Massive & $0$       & 0.0288 & 0.0574 & 0.0845 & 0.1085 & 0.7209 \\
($v_{rot}=0$) & $-1$    & 0.0311 & 0.0545 & 0.0886 & 0.1070 & 0.7188 \\
& $-2$    & 0.0351 & 0.0543 & 0.0962 & 0.1097 & 0.7048 \\
& $-3$    & 0.0374 & 0.0570 & 0.0950 & 0.1080 & 0.7026 \\
\hline
Massive & 0 & 0.0330 & 0.0573 & 0.0934 & 0.1146 & 0.7017 \\
($v_{rot}=150$) & $-1$    & 0.0627 & 0.0494 & 0.1698 & 0.1352 & 0.5829 \\
& $-2$    & 0.0526 & 0.0435 & 0.1667 & 0.1429 & 0.5943 \\
& $-3$    & 0.0291 & 0.0555 & 0.1038 & 0.2780 & 0.5337 \\
\hline
Massive & $0$       & 0.0389 & 0.0551 & 0.1170 & 0.1448 & 0.6443 \\
($v_{rot}=300$) & $-1$    & 0.0332 & 0.0273 & 0.0823 & 0.0624 & 0.7948 \\
& $-2$    & 0.0315 & 0.0150 & 0.0785 & 0.0814 & 0.7935 \\
& $-3$    & 0.0334 & 0.0283 & 0.0710 & 0.0505 & 0.8169 \\
\hline
Solar (s) & & 0.0274 & 0.0187 & 0.0892 & 0.0825 & 0.7822 \\
Solar (r) & & 0.0000 & 0.4177 & 0.0000 & 0.3341 & 0.2482 \\
\hline
\end{tabular}
\begin{tablenotes}
\item{} \textbf{Notes.} {All theoretical yields used to compute these fractions are integrated over a standard Salpeter initial mass function. Velocities are given in km~s$^{-1}$. Solar s- and r-process fractions are from \cite{Prantzos20}.} 
\end{tablenotes}
\end{threeparttable}
\end{table}
As shown in Table~\ref{tab:ratios}, the isotopic fractions from both AGB and massive stars converge toward values that are remarkably similar to the solar system $s$-process distribution. At low metallicities, slightly higher fractions of $^{138}\text{Ba}$ are found at the expense of other isotopes. Consequently, adopting standard solar fractions provides a robust approximation. 
However, within the framework of a heterogeneous Galactic chemical evolution, localized pollution from specific sources can enrich the interstellar medium with slightly different isotopic ratios. Such local heterogeneities produce deviations that solar s-process fractions fail to capture.
A striking example occurs at low metallicities, where massive stars rotating at $v = 150$~km~s$^{-1}$ exhibit a significantly enhanced fraction of $^{136,137}\mathrm{Ba}$. This enhancement is directly linked to the intense production of the unstable isotopes $^{136,137}\mathrm{Cs}$, which subsequently decay into $^{136,137}\mathrm{Ba}$. Interestingly, this localized overproduction is absent in non-rotating massive stars and is also strongly suppressed in the fastest rotators ($v = 300$~km~s$^{-1}$).

Although these localized variations can be significant, they cannot be fully accounted for in our measurement methodology. We therefore adopt the solar isotopic mixture as a practical and generally reliable approximation for our analysis.

\subsection{Barium abundance
}
\label{sec:barium}

\begin{figure}
    \centering
    \includegraphics[width=0.49\linewidth]{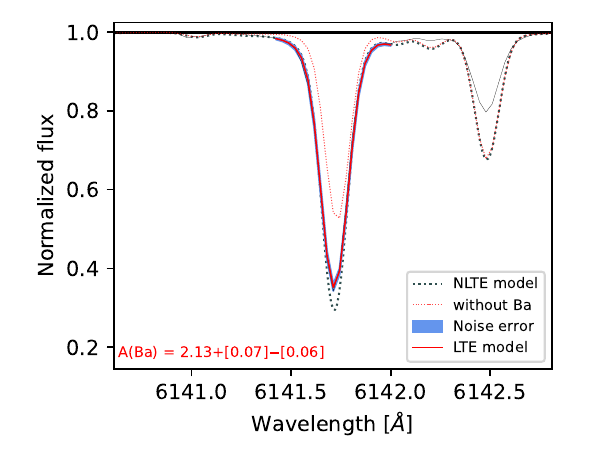}
    \includegraphics[width=0.49\linewidth]{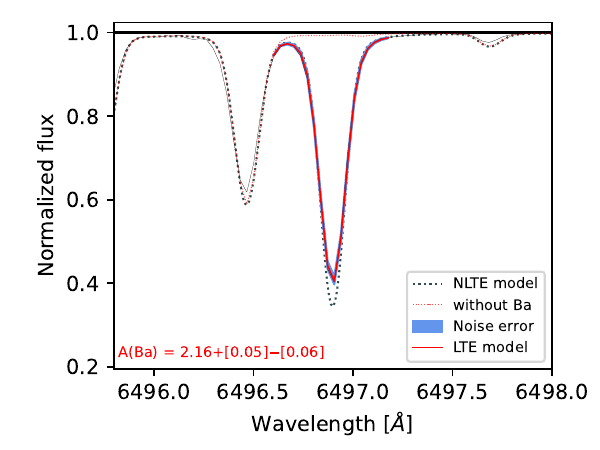}
    \caption{\tiny Line fits of the subordinate Ba lines in the solar spectrum. The observational spectrum is represented by the thin black line. Fits and corresponding A(Ba) determinations are done under 1D~LTE (red solid lines). Synthetic spectra produced under 1D~NLTE from A(Ba) determined under 1D~LTE are represented by the dark dotted lines.}
    \label{fig:sun}
\end{figure}

\begin{figure*}
    \centering
    \includegraphics[width=0.33\linewidth]{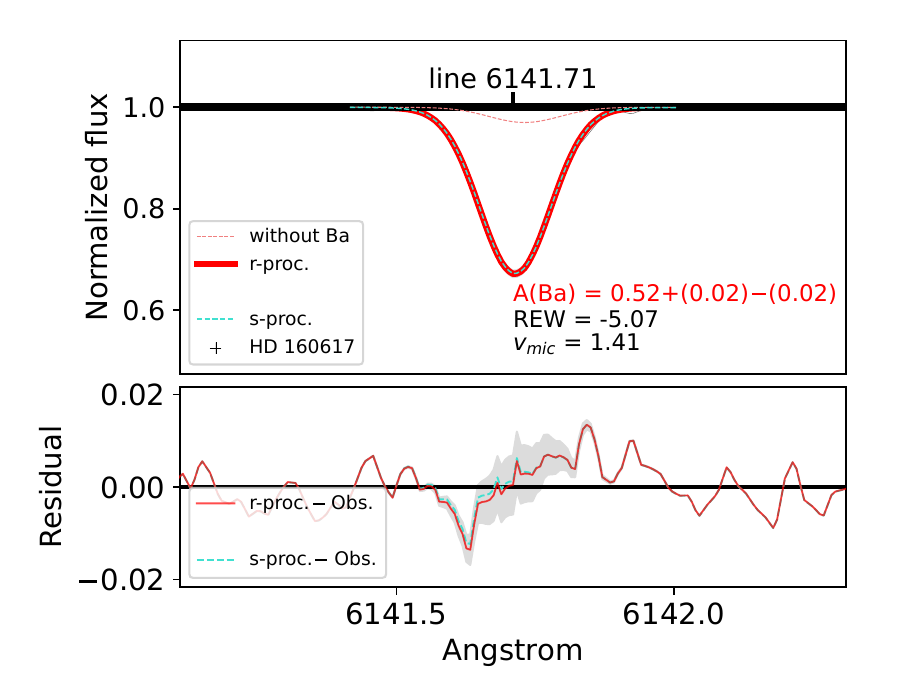}
    \includegraphics[width=0.33\linewidth]{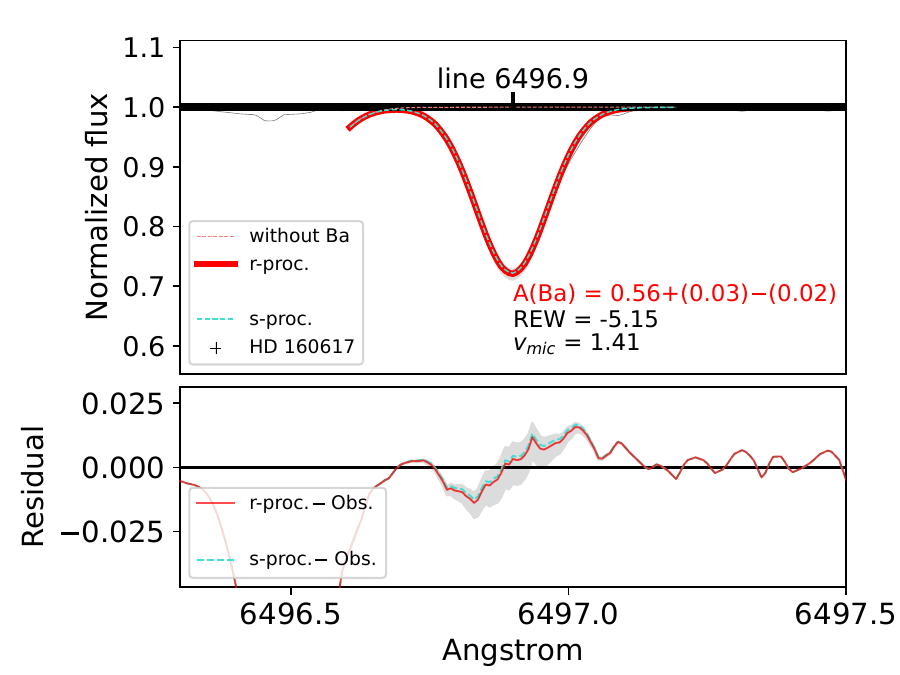}
    \includegraphics[width=0.33\linewidth]{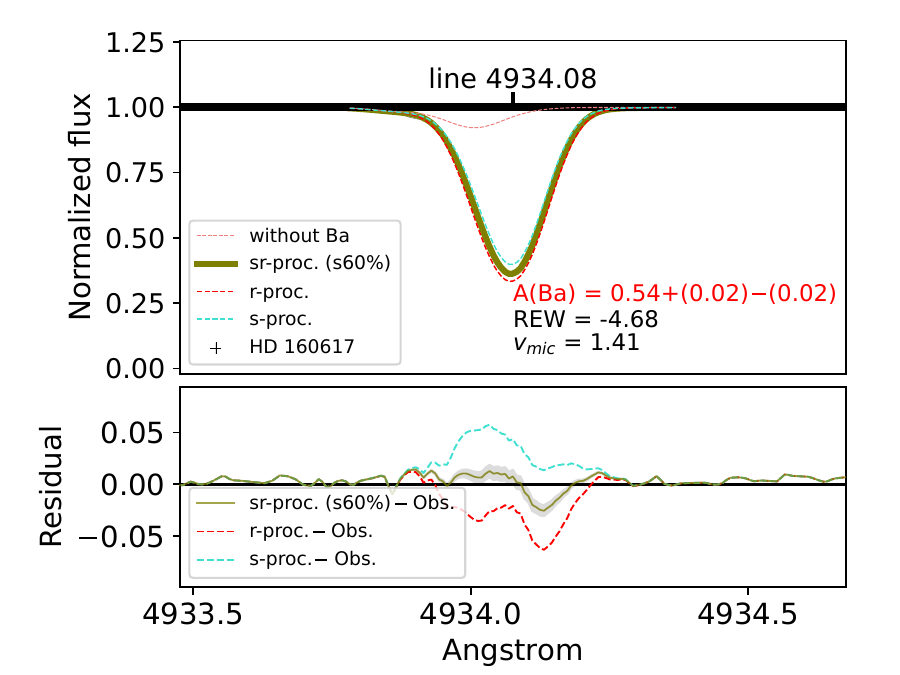}
    \caption{\tiny Line-profile fits of the three \ion{Ba}{II} lines for HD 160617. The observed UVES spectrum has $R = 51\,600$ and $S/N = 330$.
    The left and central panels show the subordinate lines, for which the best-fit abundances and their noise-related errors are indicated.
    The reduced equivalent widths (REW) of the lines and the $v_{mic}$ value are also noted.
    Synthetic profiles corresponding to s- and r-process isotopic mixtures are over-plotted using different colours,
    while the shaded gray regions in the residual panels represent the flux variation due to the noise.
    The right panel presents line-profile fits assuming the average abundance of the subordinate lines A(Ba) = 0.54~dex.
    Profiles associated with pure s-, pure r-, and a 60\% s-process mixture are shown, as indicated in the legend.
    The residuals display the flux differences between synthetic and observed profiles, with the shaded area marking the noise-induced variation, as in the left and central panels.}
    \label{fig:HD160617}
\end{figure*}

{Spectral synthesis is performed using Turbospectrum to fit the subordinate \ion{Ba}{II} lines at 6141 and 6496~\AA.
The odd isotopes $^{135}\mathrm{Ba}$ and $^{137}\mathrm{Ba}$ present hyperfine splitting of the energy levels due to their non-zero nuclear spin. We adopted the wavelengths and relative strengths of the hyperfine structure components (HFS) for the spectral lines 6141.71 and 6496.9~\AA\ from Table A.3 in \cite{Gallagher2020A&A...634A..55G}. The adopted total values of log $gf$ for the transitions at 6141.71 and 6496.9~\AA\ are $-0.032$ and $-0.407$, respectively. Here, $g = 2J+1$ is the statistical weight of the lower energy level of the transition, and $f$ is the oscillator strength. These values were taken from \cite{Gallagher2020A&A...634A..55G}, who derived the log $gf$ values from highly accurate branching fraction measurements reported by \cite{2015PhRvA..91d0501D} (for 6496.9~\AA) and \cite{2016NatSR...629772D} (for 6141.71~\AA), obtained using an ion trap experiment. These measurements were combined with the best available lifetime values to yield transition probabilities with uncertainties below 1.2\%.
Hyperfine splitting components and the corresponding log~$gf$ for each component, obtained as:
\begin{equation}
    gf_\mathrm{component} = gf_\mathrm{total} \cdot \mathrm{relative~strength}
    \label{eq:gf}
\end{equation}
are listed in Table~\ref{tab:tab1} in Appendix.}
The line synthesis requires relative contributions for each isotope from the s- and r-processes, we adopted the solar values listed in Table~\ref{tab:ratios}.

Our first line synthesis test is done with the Sun. Figure~\ref{fig:sun} shows line fits adopting the standard\footnote{IAU 2015 RESOLUTION B3 parameters [Fe/H] = 0~dex, \teff$_\odot = 5772$~K, and \logg $= 4.44$~dex  \citep{prsa2016AJ....152...41P}.} parameters.
1D~LTE synthesis yields an averaged value of A(Ba)~$= 2.15\pm 0.02$~dex, compatible with the canonical chondritic abundance A(Ba) $= 2.18 \pm 0.02$ \citep{lodders2009LanB...4B..712L}.
The figure also shows 1D~NLTE profiles computed with the A(Ba) values determined from 1D~LTE. These are systematically deeper, implying that matching the observational line strengths under 1D~NLTE requires lower abundances. We indeed obtain 1D~NLTE abundances lower by $\sim$0.16~dex relative to the 1D~LTE values for both lines, in agreement with the results of \cite{Gallagher2020A&A...634A..55G}. Remarkably, the latter authors, in their Table~3, show marginal corrections (+0.02 to +0.04~dex) when three-dimensional (3D) radiation-hydrodynamical
model atmosphere simulations under NLTE  are applied for those lines in the Sun; thereby validating the fidelity of 1D~LTE for the solar case. 3D~NLTE corrections for the entire sample are assessed in Sect.~ \ref{sec:fiducial}.

Internal uncertainties of A(Ba) were estimated from the standard deviation of individual line abundances or, when only one line is available, from the fitting error due to spectral noise. The left and central panels of Fig.~\ref{fig:HD160617} show representative fits for the dwarf star HD~160617, with line profiles corresponding to the r- and s- contributions overplotted, illustrating their marginal differences.
Total uncertainties combine the internal error and contributions from all other parameters added in quadrature. Errors in A(Ba) induced by typical uncertainties in \logg\ and [Fe/H] are negligible for dwarfs ($\lesssim 0.005$~dex), for giants only \logg\ errors may affect by 0.05~dex at most. 
The impact of \teff\ errors was estimated using the values in Table~\ref{tab:parameters} in the line synthesis procedure. With \teff\ precisions of $\sim$30~K for dwarfs and $\sim$20~K for giants, the corresponding effect on A(Ba) is $\pm$0.03 and $\pm$0.02~dex, respectively. Microturbulence has a negligible impact on the A(Ba) error budget for dwarfs ($\lesssim 0.002$~dex), whereas for giants it contributes nearly linearly with \logg\ as indicated in Table~4 of Paper~I.
Overall, the total error budget for dwarfs is dominated by internal uncertainty and \teff\ errors, contributing on average 0.03 and 0.05~dex, respectively. For giants, the dominant source of uncertainty is $v_{mic}$ ($\pm^{0.1}_{0.2}$~km~s$^{-1}$), changing A(Ba) by the same quantity in opposite sense. Surface gravity, \teff, and internal errors contribute about $\pm$0.05, $\pm$0.04, and $\pm$0.04~dex, respectively. 

The results obtained under the 1D~LTE assumption are presented in Table~\ref{tab:barium}, while the 1D~NLTE results are listed in Table~\ref{tab:barium_NLTE}. A comparison between the two sets of values shows that the 1D~NLTE abundances are systematically lower than the 1D~LTE ones. Moreover, the discrepancy increases with increasing A(Ba), as shown in Fig.~\ref{fig:Ba_NLTE}.
In Sect.~\ref{sec:fiducial}, we present our assessment of which set of values most closely approximates the true measurements.

\subsection{Barium isotopic ratios}
\label{sec:isotopes}

The Ba isotopic ratios were determined  by fitting the profile of the 4934~\AA\ resonance line, with the A(Ba) abundances from Table~\ref{tab:barium} held fixed. 
The line synthesis incorporates the {solar theoretical isotopic mixtures given in Table~\ref{tab:ratios}, as justified in Sect.~\ref{sec:iso_theoric}.} 
Because of the hyperfine splitting, the isotopes $^{135}$Ba and $^{137}$Ba are shifted out of the line centre (4934.076~\AA) towards the wings. 
Since the r-process produces a higher fraction of these isotopes, its line profiles are broader and shallower than those of the s-process. This is evident in spectra with extremely high resolution ($R \sim 500\,000$, see the simulated spectra in Fig.~2 of Paper~I). At high to moderate resolutions ($20\,000 \lesssim R \lesssim 200\,000$), the r-process profiles primarily appear stronger than the s-process ones. However, the flux difference between them in each wavelength bin decreases with decreasing resolution, making the signal-to-noise ratio increasingly critical.

A detailed description of our line profile fitting procedure and the adopted hyperfine-structure data can be found in Paper~I. Briefly, our strategy consists on assuming that the isotopic composition that forms the Ba line of a star can be represented as a complementary combination of s- and r-process contributions. Accordingly, we express our results in terms of the s-process fraction, ranging from 0\% (pure r-process) to 100\% (pure s-process).
The right panel of Fig.~\ref{fig:HD160617} illustrates the sensitivity of the 4934~\AA\ resonance line to variations in the isotopic mixture.
The residuals show that the synthetic profiles corresponding to pure s- and pure r-process isotopic compositions remain well outside the shaded noise interval, highlighting the robustness of the inferred s-process fraction.
The resulting s-process contributions, and the corresponding $f_{\rm odd} \equiv [N(^{135} \rm Ba) + N(^{137} \rm Ba)]/N(\rm Ba)$ fractions, for all stars are listed in Table~\ref{tab:barium}.

In giant stars with metallicities around [Fe/H]~$\sim -1.5$ or higher, the 4934~\AA\ \ion{Ba}{II} resonance line is significantly blended with a Fe feature on its blue wing (see Fig.~9 in Paper~I). This contamination produces an over-saturated line profile that responds inconsistently to 1D modelling, potentially yielding unreliable isotopic ratios. Paper~I provides an empirical calibration relating the equivalent width of the 4934~\AA\ line to an adjusted microturbulence parameter. Within our sample, only the giant HD~45282 falls within this metallicity regime, and for this star we adopt the calibrated microturbulence. For all other giants, which are considerably more metal-poor ([Fe/H] $< -2$~dex), the Fe blend is negligible, and the specialised microturbulence calibration is therefore unnecessary.

Uncertainties in the isotopic ratios—expressed as percentages of the s-process contribution—arising from the uncertainties of the atmospheric parameters are adopted from Table~6 of Paper~I, and are propagated into the errors reported in Table~\ref{tab:barium}.
Effects from the parameters are listed separately for the giants.
The $v_{mic}$ uncertainty also impacts the isotopic fractions via A(Ba) variations.
In dwarfs this effect is negligible because the variations of both parameters  cancel each other.
In giants, the former is determined with a precision of $\pm^{0.1}_{0.2}$~km~s$^{-1}$, which due to a nearly one-to-one relation induces an A(Ba) uncertainty of $\pm^{0.2}_{0.1}$~dex.
Since these two quantities affect the isotopic fractions in opposite directions, we include in the uncertainty budget the simultaneous variation of both during the line-profile fitting. 
Section.~\ref{sec:NLTE} shows that this simultaneous variation strongly affects isotopic fraction determinations of giant stars with resonance lines of equivalent width (EW) lower than $\sim$140~m\AA\ (see Fig.~\ref{fig:EW_sproc}). For this reason, the EW is deemed the main data limitation for the application of the method in Sect.~\ref{sec:limitations}.
To include the sole effect of the A(Ba) uncertainty, 
we add in the budget the effect of its typical internal error ($\pm0.04$~dex), which corresponds to 11\%.

\subsection{1D~LTE versus 1D~NLTE treatment of barium}
\label{sec:fiducial}

\begin{figure}[!htbp]
    \centering
    \includegraphics[width=0.9\linewidth]{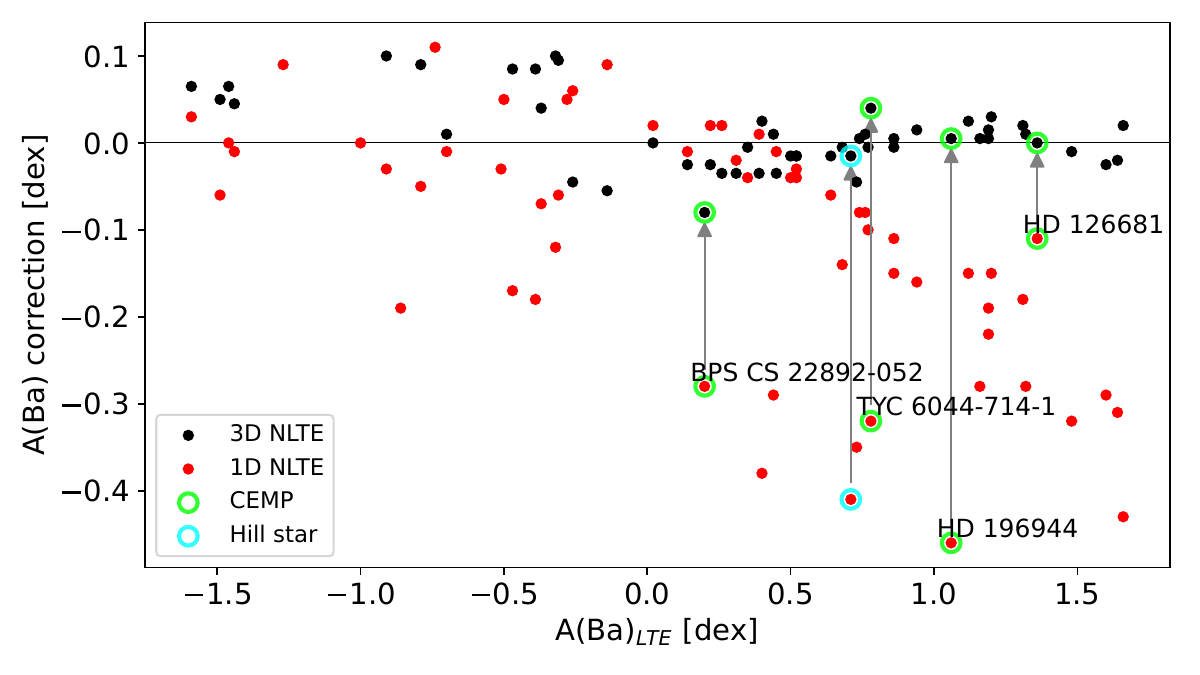}
    \caption{\tiny 3D~NLTE and 1D~NLTE corrections to 1D~LTE A(Ba) abundances as function of 1D~LTE A(Ba). Averaged corrections for both the 6141 and 6496~\AA\ lines are shown. CEMP and Ba-enhanced stars are identified as indicated in the legend. Arrows connect the 1D NLTE and 3D NLTE corrections for the same peculiar stars.}
    \label{fig:3DNLTE_cor}
\end{figure}

We assess the accuracy of 1D~LTE and 1D~NLTE A(Ba) determinations adopting the 3D~NLTE scale as the standard.
We interpolated 3D NLTE corrections\footnote{Available at \url{https://www.chetec-infra.eu/3dnlte/abundance-corrections/barium/}.} computed by the code Linfor3D \citep{steffen2013MSAIS..24...37S} based on the CO$^5$BOLD model atmosphere grids \citep{freytag2012JCoPh.231..919F, ludwig2009MmSAI..80..711L} and the model atom in \cite{Gallagher2020A&A...634A..55G}.
Figure~\ref{fig:3DNLTE_cor} shows that the 3D NLTE corrections are lower than 0.11~dex in absolute value for all stars in the sample. The largest corrections are found for giants with lower A(Ba). The Sneden star exhibits small negative corrections; this behaviour arises from the interplay between its atmospheric parameters and the equivalent widths of its subordinate lines. This is shown in Fig.~\ref{fig:3Dcors}, where its correction grid is interpolated.
Figure~\ref{fig:3DNLTE_cor} also includes the 1D NLTE corrections discussed in Sect.~\ref{sec:NLTE}. 
These are generally negative and imply a significant reduction in the derived abundances. They affect isotopic ratios biasing the diagnoses towards r-process dominance, as explained in detail in Sect.~\ref{sec:NLTE}. By contrast, the modest amplitude of the 3D~NLTE corrections indicates that 1D~LTE abundances already provide a close approximation to the true values.
In most cases 3D~NLTE corrections to A(Ba) from subordinate and resonance lines compensate or are negligible (e.g. $\lesssim \pm 0.05$). However, in rare cases, the corrections may accumulate inducing significant impact in isotopic fraction determinations. A single case appears in our sample for BD$-18\,5550$ with a weak resonance line (EW $\sim 78$~m\AA), see Sect.~\ref{sec:BD18}.

\begin{figure*}[!htbp]
    \centering
    \includegraphics[width=0.7\linewidth]{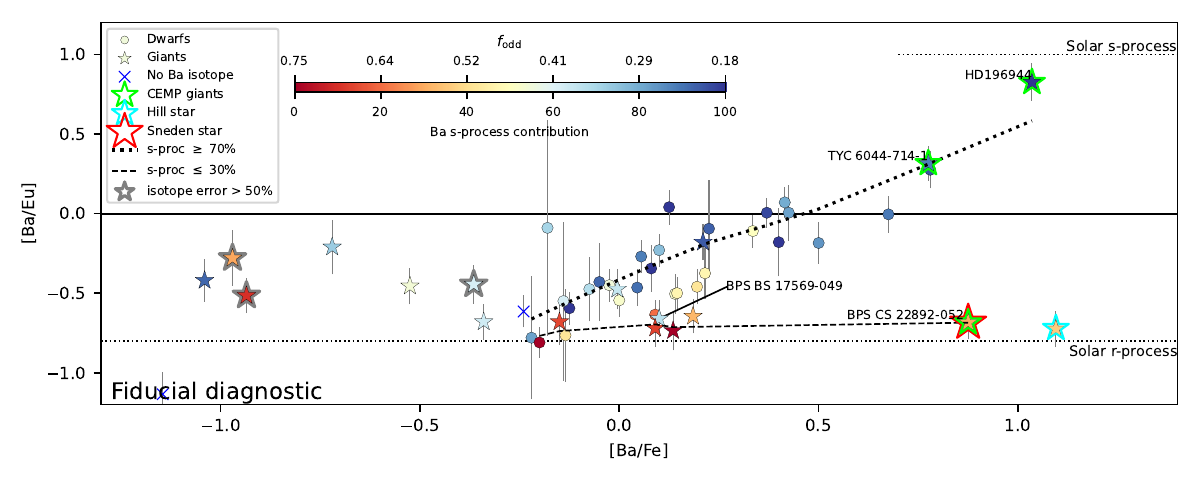}
    \includegraphics[width=0.7\linewidth]{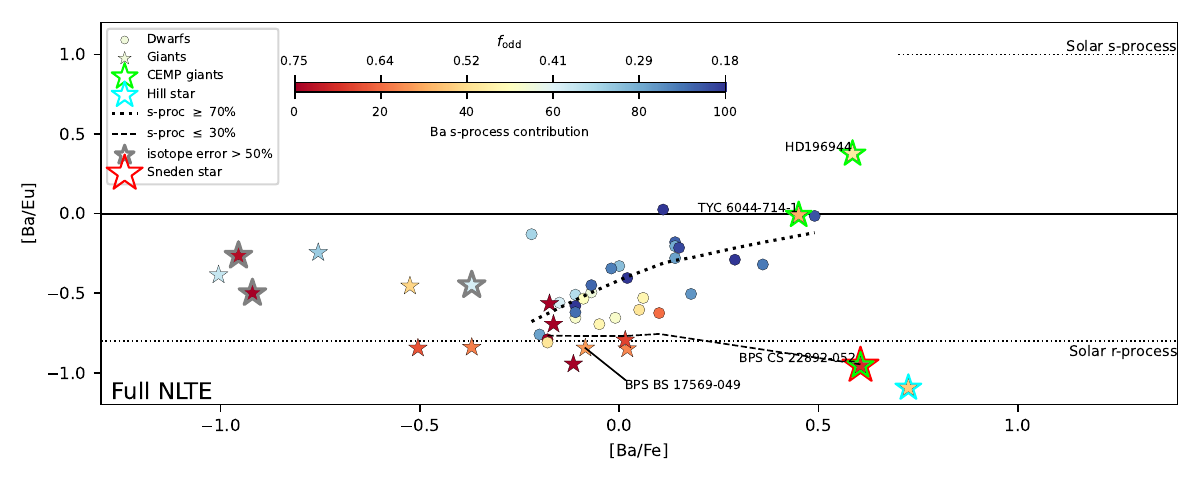}
    \caption{\tiny [Ba/Eu] versus [Ba/Fe] diagram. Dwarf and giant stars are indicated as specified in the legend. CEMP stars are highlighted by green. The Hill star (BPS~CS~31082-001) and Sneden star (BPS~CS~22892-052) are labelled. Stars with s-process contributions with uncertainties larger than 50\% are marked by gray contours. The fractional s-process contribution percentage is colour-coded according to the bar, it includes the F$_{odd}$ normalised fraction scale.  The pure solar s- and r-process reference values (1.0 and $-0.8$~dex, respectively; \citealt{Simmerer2004ApJ...617.1091S}) are shown as dotted horizontal lines.  A LOWESS regression fitted to stars with s-process contributions greater than 70\% is shown as a thick dotted line, while one regression fitted to stars with s-process contributions lower than 30\% is represented by a dashed line. The top panel presents the fiducial diagnostic, adopting 1D~LTE A(Ba) abundances and isotopic ratios from 1D~NLTE line synthesis. The bottom panel shows the full 1D~NLTE diagnostic, where both A(Ba) and isotopic ratios are determined from 1D~NLTE calculations.}
    \label{fig:BaEu_BaFe}
\end{figure*}

Figure~\ref{fig:BaEu_BaFe} displays [Ba/Eu]\footnote{Element fractions adopt the canonical solar values of \citet{asplund2009ARA&A..47..481A} as reference.}–[Ba/Fe] diagrams for our fiducial diagnostic (A(Ba) from 1D~LTE and isotopic fraction from 1D~NLTE) and full 1D~NLTE modelling.
Full 1D~LTE modelling is shown to yield unreliable results in Sect.~\ref{sec:NLTE}.
In principle, we expect giant and dwarf stars to follow the same path in the [Ba/Eu]–[Ba/Fe] plane, with the Ba percentage from the s-process increasing as both [Ba/Eu] and [Ba/Fe] rise.
In the upper panel, where we adopt the 1D~LTE approach for A(Ba) and 1D~NLTE for the isotopic ratios, we see that stars with an s-process contribution of at least 70\% form a tight sequence in which [Ba/Eu] correlates with [Ba/Fe]. This sequence delineates an apparent upper envelope that converges with the path of stars dominated by the r-process at [Ba/Fe] $\approx -0.25$~dex. 
With the exception of a single outlier (BPS BS 17569-049 indicated in the plot), the inferred s-process fractions of all giants —including the CEMP stars— are fully consistent with the trend defined by the dwarfs (dotted line).
In addition, all giants with [Ba/Eu] $\equiv -0.8$~dex are consistent with a major dominance of the r-process, including the standards Hill and Sneden stars (see details in Sect.~\ref{sec:peculiars}).

The lower panel, based on the full 1D~NLTE diagnostic, yields a markedly different interpretation: the r-process appears to dominate the chemistry of most giants, as explained in Sect.~\ref{sec:NLTE}, although this result is physically implausible. Moreover, the two CEMP giants located along the sequence defined by the s-process dwarfs show an inferred s-process contribution of $\sim$50\%, despite their [Ba/Eu] and [Ba/Fe] ratios indicating s-process dominance. This discrepancy could be interpreted as evidence of a partial contribution from the i- process in their atmospheres (e.g. the CEMP star HD~196944 discussed in Sect.~\ref{sec:hd196944}), given that the line profile models of the r- and i-process are nearly indistinguishable (e.g.  Fig.~2 in Paper~I).

These considerations support the adoption of our fiducial hybrid approach for measuring barium abundance and isotopic ratios. It demonstrates that the choice of 1D~LTE and 1D~NLTE significantly affects the derived isotopic ratios in giant stars. Consequently, using an inappropriate approach could lead to incorrect conclusions about barium nucleosynthesis.

\subsection{Applicability limits}
\label{sec:limitations}

\begin{figure}
    \centering
    \includegraphics[width=1\linewidth]{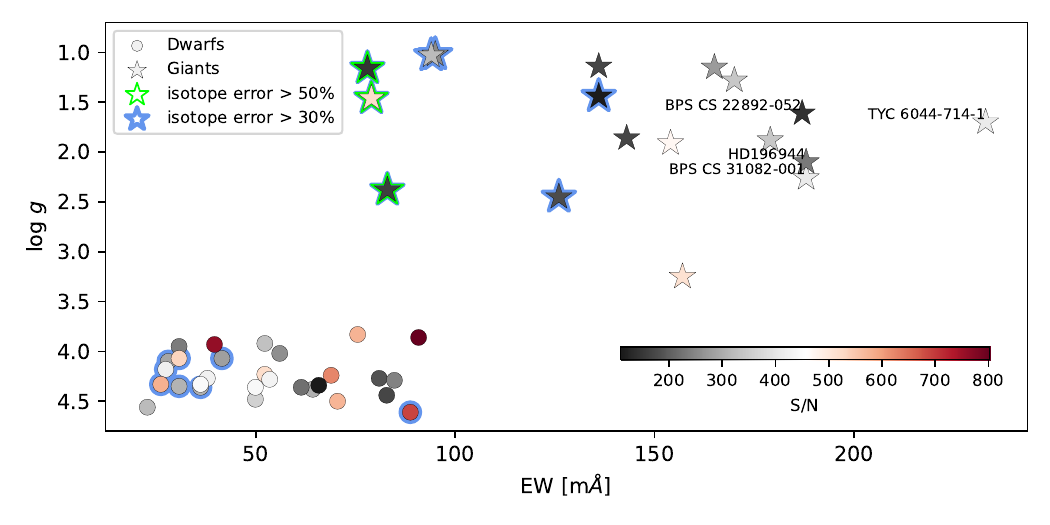}
    \caption{\tiny  Equivalent width versus surface gravity. For dwarfs, the EW  correspond to the subordinate line at 6141\AA. For the giants, the EW corresponds to the resonance line at 4934~\AA.
    The color-code indicates the S/N in Table~\ref{tab:parameters}. Isotopic fractions with precisions worse than 30\% and 50\% are marked in blue and green respectively.
    Labels indicate stars enhanced in Ba.}
    \label{fig:applicability}
\end{figure}

Throughout this section we have shown that the reliability of barium isotopic diagnostics depends on several factors, including spectral quality, line strength, and stellar parameters. The most favourable conditions are nevertheless found within a restricted region of parameter space.
For dwarfs (with $5600 \lesssim$ \teff\ $\lesssim 6700$~K) observed at the typical resolution of our sample ($R \sim 50\,000$), the method is primarily limited by the detectability of the subordinate lines. As a consequence, the applicability range is restricted to approximately [Fe/H] $\gtrsim -2$~dex and [Ba/Fe] $\gtrsim 0$ \citep[Figs.~5 and 6 of][]{giribaldi2026arXiv260511074G}, beyond which the subordinate lines become too weak to measure. Figure~\ref{fig:applicability} shows that equivalent widths of subordinate lines of at least $\sim$20~m\AA\ and spectra with S/N $\gtrsim 300$ are generally required. Under these conditions, the expected precision of the isotopic fraction (represented by the s-process percentage contribution) is typically between 30\% and 50\%, while subordinate lines stronger than $\sim$50~m\AA\ yield significantly improved constraints.

Giants (with $4600 \lesssim$ \teff\ $\lesssim 5600$~K) extend the applicability of the method down to [Fe/H] $\sim -3$~dex for chemically normal stars \citep[Figs.~5 and 6 of][]{giribaldi2026arXiv260511074G}, owing to their intrinsically stronger Ba lines. In this regime, the 4934~\AA\ resonance line becomes particularly sensitive to variations in A(Ba) for EW~$\lesssim 150$~m\AA\ (see also Sect.~\ref{sec:NLTE}). Since the EW of this line is paramount in giant stars, Fig.~\ref{fig:applicability} adopts it width as a practical proxy. In general, spectra with EW~$ \lesssim 100$~m\AA\ provide isotopic fractions with uncertainties exceeding 50\%, even at high S/N. For $100 \lesssim \rm{EW} \lesssim 150$~m\AA, precisions of about 30--50\% are achieved for spectra with S/N $\sim 200$, while stronger lines permit robust determinations already at S/N $\gtrsim 150$.
The whole set of isotopic fractions and their precisions, the atmospheric parameters, and the S/N used in this analysis are listed in Tables~\ref{tab:barium} and \ref{tab:parameters}.

\section{Galactic chemical evolution modelling}
\label{sec:discusion}

To test our observationally based results from a theoretical perspective, we adopt stochastic chemical evolution models of the early Milky Way (MW) and the GES.
The adopted models resemble the scheme adopted in the inhomogeneous/stochastic models presented in \citet{Cescutti08,Cescutti10} and later used in other works to model the evolution of the Galactic halo. The framework also successfully reproduce the physical features and metal-poor abundance patterns observed in satellite galaxies in the Local Group, as shown in e.g. \citet{Santarelli26}. Therefore, our adopted framework can be safely applied to describe the early chemical evolution in different systems. 
We refer to \citet{Santarelli26} (see their Section 7 and C) for a full description of the framework principles and equations. Below we provide a summary of the adopted model features.

\subsection{Stochastic chemical evolution model (CEM) framework}

In our models, stochasticity is introduced by dividing the galaxy into isolated cubic regions, each containing the typical mass of gas swept by a CC-SNe (see \citealt{Cescutti08}).
Within each cubic region, the model considers at each step gas inflows and outflows, gas recycling due to evolving stars and star formation. For the latter, masses of the stars formed are randomly extracted weighting them according to the initial mass function in \citet{Scalo86}. It is worth noting that to ensure good statistical results, the number of assumed volumes for the model is set to 100, each of them run for a time-interval of 2 Gyr.

To describe the different histories of star formation (SFHs) in the early MW and GES progenitor, we adopt different parameters for the two galaxies. In particular, for the MW model we set the star formation efficiency (SFE, namely the star formation rate per unit gas mass- $\nu$)  to 2 Gyr$^{-1}$ and the gas infall timescale ($\tau_{inf}$) to  0.25 Gyr, whereas for GES we adopted $\nu$=0.5 Gyr$^{-1}$ and $\tau_{inf}$=1 Gyr. For both the models, we assume outflows with a wind mass loading factor of 4 (see \citealt{Santarelli26}). The parameters for the MW model are analogues to those routinely assumed to model the in-situ halo \citep[e.g.][]{Spitoni16}. The slower rate of star formation assumed in the GES model agrees with estimates from one-zone chemical evolution models \citep{vincenzo2019MNRAS.487L..47V} as well as with the lower stellar mass of GES progenitor relative to the proto-Galaxy, given the theoretical scaling relations between galactic stellar mass and star formation efficiency \citep{Prantzos08,Komiya16}.
Nonetheless, we do not intend to fine-tune the model parameters to a specific scaling relation, due to the  uncertain mass of the GES progenitor relative to the Galaxy \citep[e.g.][]{helmi2018,mackereth20,Lane23,Carrillo23}.

Focusing on the stellar nucleosynthesis, the model includes chemical production from all the classical stellar sources (low-intermediate mass stars, massive stars, Type Ia SNe), implementing well-tested stellar yields from the literature 
(\citealt{cristallo2015ApJS..219...40C}, and references therein; \citealt{Limongi18,Iwa99}).
To trace the evolution of neutron-capture elements, 
s-process yields in AGB stars are adopted from the FRUITY database \citep[][and references therein]{cristallo2015ApJS..219...40C}.
Slow process yields from rotating massive stars yields are adopted from \citet[][their set R]{Limongi18}, for which we assume a initial rotational velocity distribution with Gaussian shape, centred at $v_{rot}=121.5$ km s$^{-1}$ and with $\sigma=114.5$ km s$^{-1}$ (see Sect. \ref{sss:model_res}). 
R-process material is assumed to be produced by Magneto-Rotationally-Driven (MRD) SNe and compact binary mergers (referred as NSM for convenience). For MRD-SNe, we adopt the yields of \citet[][their model L0.75]{Nishimura17} and a probability of such events in the progenitor mass range of 0.1. For NSM, we adopt the same yields as in \citet{Molero23}, which are derived by scaling to the solar r-process pattern from the Sr measurement in the Kilonova AT2017gfo \citep{Watson19}. We set a probability of a NSM event in the progenitor mass range of 0.08 (as derived in \citealt{Palla25}) and a fixed coalescence timescale of 150 Myr. 
As variations in r-process production are expected between progenitors of the same class, but no sampling in mass and metallicity is available \citep[see][for discussion on this point]{Molero25,Palla25}, we also assume a non-constant r-process production per event. 
Similarly to \citet{Cescutti14}, to preserve the total r-process production budget of the adopted yields for MRD-SNe and NSMs, the ejected r-process mass for an $n$ event follows this equation:
\begin{equation}
    M_{r}(n) = M_{r,0} \ (0.1+1.9 \cdot {\rm Rand}(n) ),
    \label{eq:rproc_scatter}
\end{equation}
where ${\rm Rand}(n)$ is a uniform random distribution in the rage $[0,1]$. In this way, we allow variations between 10 and $\sim$200\% relative to the yields listed above.

\subsection{Model results: r-/s-process balance and impact of massive rotators}
\label{sss:model_res}

\begin{figure*}
    \centering
    \includegraphics[width=0.8\linewidth]{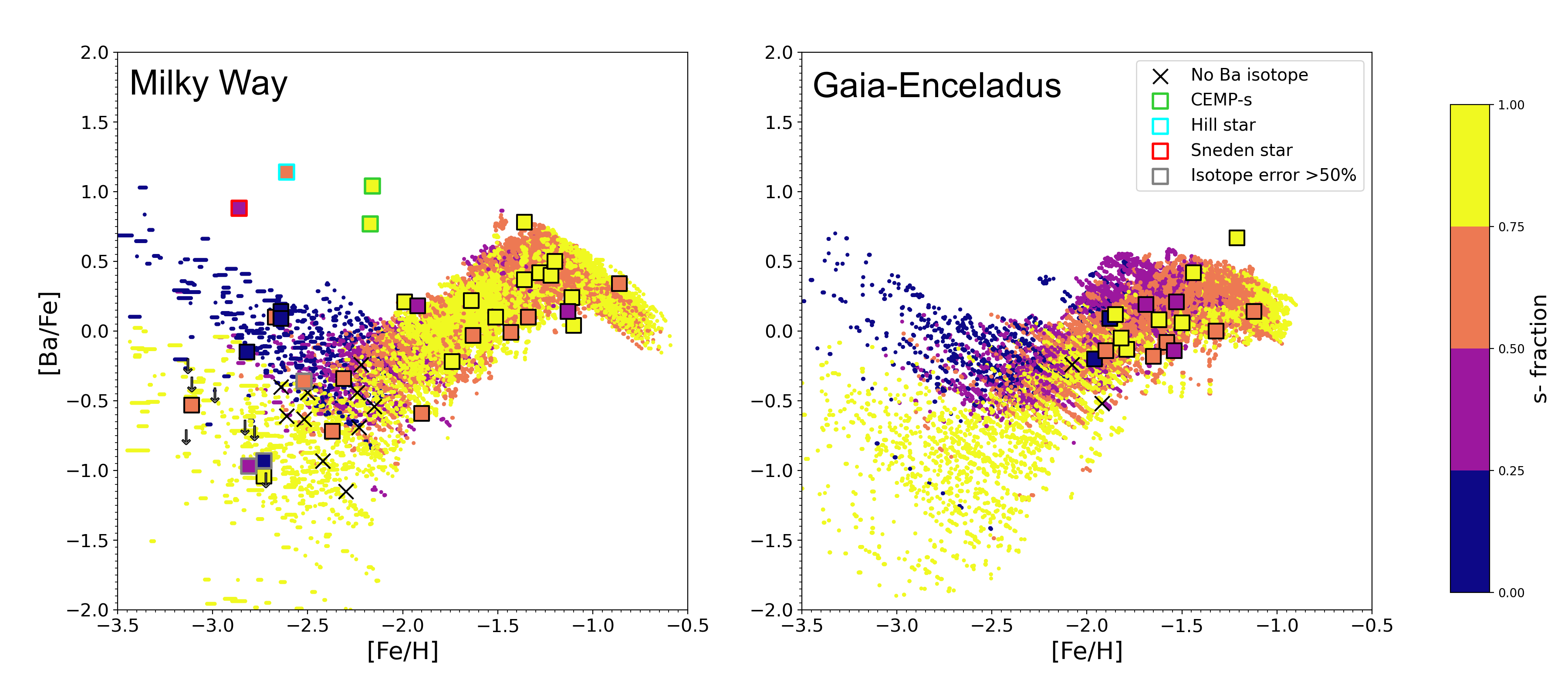}
    \caption{\tiny
    [Ba/Fe] versus [Fe/H] diagram as predicted by our fiducial stochastic chemical evolution model framework. Left panel shows predictions by the MW model, right panel for the GES model. For both panels, the colour map displays the fractional s-process contribution to Ba abundance.  Stars without s-process contribution determination are shown as black crosses, while those with s-process contributions with uncertainties larger than 50\% are marked by gray contours. CEMP-s stars, Hill star and Sneden star are highlighted by green, cyan and red contours, respectively. Upper limits are displayed with grey arrows.
    }
    \label{fig:BaFe_sfract}
\end{figure*}

In this section, we show the outcomes of the models for early MW and GES evolution in comparison with the sample presented in the previous Sections.
This comparison is of particular importance to extract further information and progress in the understanding of the origin of neutron-capture elements. 
Indeed, in this work we have access not only to usual chemical abundance diagrams, but also to information as the s- and r-process contributions in stars.
{Theoretical CEM predictions for this diagnostic were already presented by \citep{Cescutti14}; however, the absence of suitable observational constraints prevented its use in chemical evolution analyses until very recently, when studies such as \citet{Sitnova25, sitnova2025A&A...704A.103S} finally began to investigate its diagnostic potential.}

\paragraph{The balance between r- and s- process in Ba production}

Figure~\ref{fig:BaFe_sfract} shows the evolution of [Ba/Fe] as function of [Fe/H] for the MW model (left panel) and the GES model (right panel). 
In both panels, observational data and model outcomes are colour-coded according to the Ba s-process contribution.
The theoretical framework well-reproduce the patterns shown by the data in the two panels of Fig. \ref{fig:BaFe_sfract}, with the observed abundance trends and spreads in [Ba/Fe] for a given [Fe/H] being genuinely predicted by the two models.  The only exceptions are represented by i) the  Sneden and Hill stars and ii) Ross~892 with 
[Ba/Fe]~$\sim 0.7$ dex at
[Fe/H] $\sim -1.2$ dex in the right panel\footnote{We do not consider in the comparison the labelled CEMP-s stars, as the extrinsic origin (AGB binary companion enrichment) of their s-process abundances cannot be accounted within a CEM framework.}. For the former, the extreme enhancement in all the neutron-capture elements at their metallicity is a well-known problem in chemical evolution studies. Regarding the GES star Ross 892, a previous study determined a relative young age of $\sim 9$ Gyr for this star \citep{giribaldi2023A&A...673A..18G}. Therefore, chemical enrichment at such age cannot be properly accounted in our modelling framework, which is limited to 2 Gyr of evolution.

The MW and GES models also capture the observed behaviour of the s-process fraction: this is a non-uniform trend as function of [Ba/Fe] and [Fe/H], {with slightly different behaviour depending on the progenitor galaxy. Indeed, GES (Figure \ref{fig:BaFe_sfract} right panel) shows on average lower s-process fractions than the MW (Figure \ref{fig:BaFe_sfract} left panel), especially for [Fe/H]$>-2$ dex: this indicates a larger relative contribution by delayed r-process sources as NSMs, stemming from the lower rate of chemical enrichment typical of a dwarf system. 
Despite this difference, both the systems display a common scheme. In the metal-poor tail ([Fe/H]$\lesssim-2.5$ dex) and roughly over-solar [Ba/Fe], the r-process is the dominant source of Ba, as observational values suggest. 
Still at [Fe/H]~$\lesssim-2.5$ dex, but at sub-solar [Ba/Fe], s-process dominates the Ba production. However, low s- process fractions are still theoretically possible (see non-yellow dots in the bottom left part of the Figure panels), allowing to explain the two stars with low ($<50\%$) s-fraction even without accounting for possible biases in Ba abundance and s-fraction determination (see Appendix \ref{sec:BD18}).
In the other regions of the abundance diagram, instead, a generalised, significant scatter in the Ba s-process fraction is instead seen at a given [Ba/Fe] vs. [Fe/H] location.}
To help in the visualisation of this scatter, Figure \ref{fig:BaFe_sfract} shows the predictions for individual time steps in each stochastic volume. It is clearly observed that radically different s-process fractions (even more than 50$\%$ difference) can be produced at a given [Ba/Fe] and [Fe/H] point, in line with the observational data.
Indeed, by comparing the s-fraction differences between model predictions and the observational data at the different locations in the abundance plane, we find median differences smaller than $\pm 10\%$ for both the proto-MW and GES.

\begin{figure}
    \centering
    \includegraphics[width=0.85\linewidth]{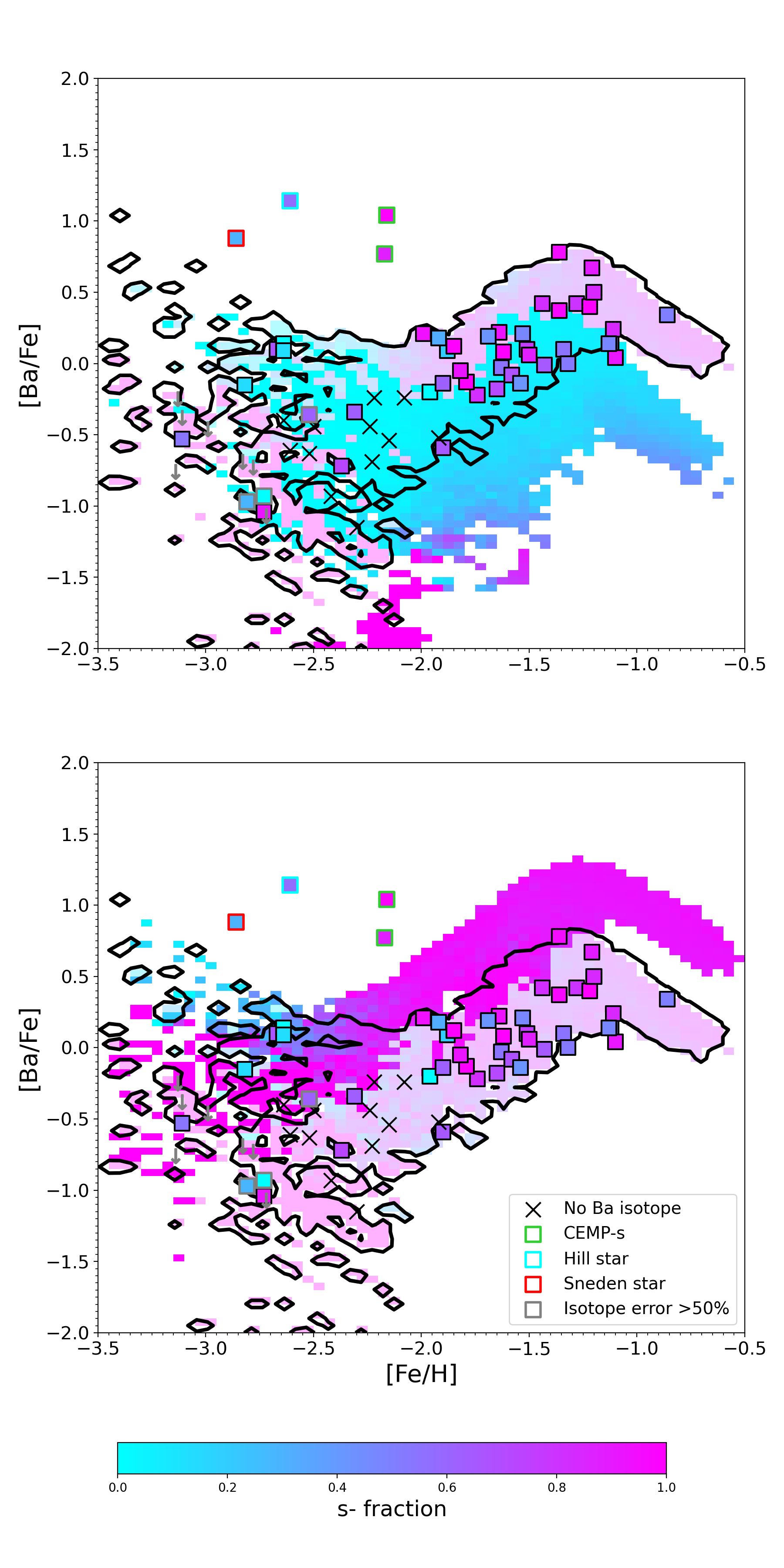}
    \caption{\tiny [Ba/Fe] versus [Fe/H] diagram as predicted by stochastic chemical evolution models adopting different yield prescriptions for massive stars than the fiducial model. Top panel compares the fiducial predictions with a model adopting an {initial rotational velocity (IRV) distribution skewed towards lower velocities ($v_{rot}$ uniform in range 0-150 km s$^{-1}$)}. Bottom panel compares fiducial predictions with a model adopting an IRV skewed towards higher velocities ($v_{rot}$ uniform in range 150-300 km s$^{-1}$). For both panels, black contours represent the predictions by the fiducial model, while the colour map displays the fractional s-process contribution to Ba abundance as predicted by the model with alternative yield prescriptions. Data legend is as in Fig. \ref{fig:BaFe_sfract}.    
    }
    \label{fig:BaFe_yields}
\end{figure}

\paragraph{The impact of massive stars in Ba production at low metallicity}

In addition, the [Ba/Fe] vs. [Fe/H] diagram  represents a useful discriminant on the yields prescriptions adopted for massive stars: the impact of AGB stars on the Ba production is negligible for metallicities [Fe/H]$\lesssim$-1.5 dex, as they are reached within timescales smaller than the lifetimes of AGBs with a significant net Ba production (see also Fig. \ref{fig:BaFe_AGB} in Appendix). Moreover, the r-process component is constrained by the [Eu/Fe] vs. [Fe/H] diagram (being Eu an almost pure r-process element, e.g. \citealt{Sneden2008ARA&A..46..241S}), which general trend is reproduced by the models (see Fig. \ref{fig:EuFe} in Appendix).
For these reasons, we can use [Ba/Fe] vs. [Fe/H] to get insights on the initial rotational velocity (IRV) distribution of massive stars at low metallicity.
In Figure \ref{fig:BaFe_yields} we compare the prediction for three MW models with identical physical parameters but different recipes for the IRV distribution of massive stars. Together with our fiducial setup, we display the results for distributions skewed towards lower 
(uniform distribution between 0 and 150 km s$^{-1}$) and larger $v_{rot}$ (uniform distribution between 150 and 300 km s$^{-1}$).
On the one hand, the model with lower characteristic $v_{rot}$ (Figure \ref{fig:BaFe_yields} top panel) shows lower [Ba/Fe] ratios than observations, with an average low s-process fraction (tending to 0) for metallicities [Fe/H]$\gtrsim-2$ dex. Conversely, the model with larger characteristic $v_{rot}$ (Fig. \ref{fig:BaFe_yields} bottom panel) shows an excess of Ba for a given metallicity, together with an average s-process contribution tending to 100\% in almost all locations of the abundance diagram. 
This comparison between these different predicted patterns tells us about the relative balance between s- and r-process sources in Ba production in the structures within the Galactic halo, which is very sensitive to the IRV distribution adopted for massive stars. 
In addition, it supports the presence of a non-negligible component of mildly rotating massive stellar population at least up to moderate metallicities ([Fe/H]$\lesssim-1$ dex), in line with \citet{Prantzos20} but at variance with other studies (\citealt{Rizzuti21} but using the \citealt{Limongi18} set F\footnote{Relative to the set R adopted in this paper, set F adopts different recipes in the choice of the remnant mass cut (set R including mixing and fallback, set F not including the mechanism) and the maximum mass for successful SN explosion ($m<30\ \rm M_\odot$ for set R, no limit for set F).}; \citealt{Molero24}). {Nonetheless, drawing further considerations on this matter goes beyond the scope of this study, as i) the relatively low data statistics available and ii) the dependence on s-process yields on other mechanism than stellar rotation (e.g. mixing and fallback mechanism, limiting initial stellar mass cut for failed SN explosions, see \citealt{Limongi18}) prevent more quantitative characterisation on the IRV distribution evolution.}

\paragraph{The origin of the [Ba/Eu] vs. [Ba/Fe] sequence} 

The results shown in Fig.~\ref{fig:BaEu_BaFe} indicate that almost all stars with a significant s-process contribution ($\gtrsim 70\%$) in Ba form a tight sequence in the [Ba/Eu] versus [Ba/Fe] plane, starting at [Ba/Fe] $\simeq -0.25$ dex and [Ba/Eu] $\simeq -0.6$ dex. This correlation for stars with high s-process fractions suggests that the diagram can serve as a discriminant between r- and s-process-enriched stars in the metal-poor regime. Accordingly, we examined the predictions of our model framework in this diagram to gain further theoretical insight into the origin of the observed [Ba/Eu]–[Ba/Fe] sequence.

\begin{figure*}
    \centering
    \includegraphics[width=0.85\linewidth]{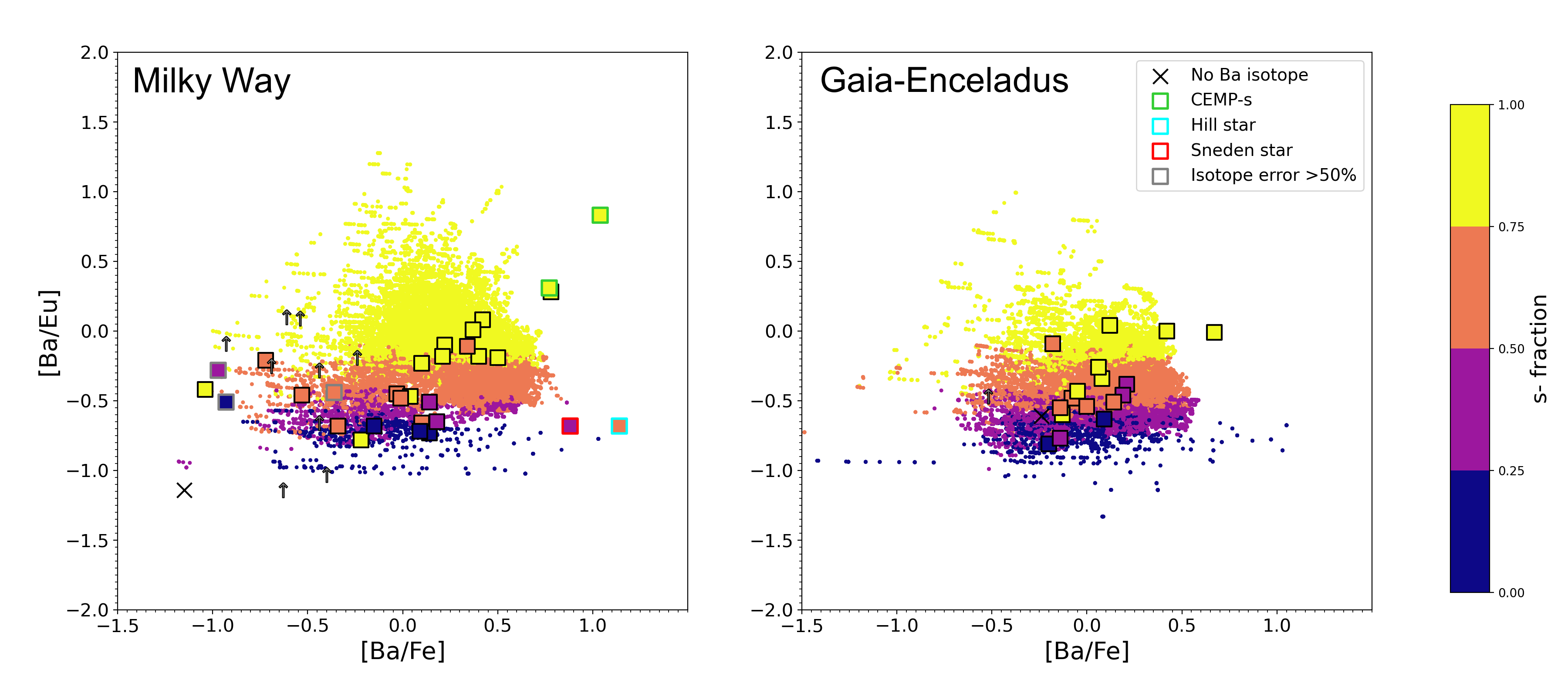}
    \caption{\tiny [Ba/Eu] versus [Ba/Fe] diagram as predicted by our fiducial stochastic chemical evolution model framework. Left panel shows predictions by the MW model, right panel for the GES model. For both panels, the colormap displays the fractional s-process contribution to Ba abundance. Data legend is as in Fig. \ref{fig:BaFe_sfract}, except for lower limits displayed with grey arrows.
    }
    \label{fig:BaEu_BaFe_sfract}
\end{figure*}

The results of the comparison are shown in Fig.~\ref{fig:BaEu_BaFe_sfract}. Despite some differences in the predicted patterns driven by the different star formation histories of the MW and GES models, both panels indicate that the [Ba/Eu]–[Ba/Fe] plane does not show a clear trend for identifying s-process-enriched stars. Instead, the predicted Ba s-fraction points to [Ba/Eu] as the primary tracer of the s-process contribution to stellar abundances. This is also illustrated in Fig.~\ref{fig:BaEu_FeH_sfract} in the Appendix, where the [Ba/Eu] versus [Fe/H] diagram shows low s-fractions smoothly transitioning to high s-fractions with increasing [Ba/Eu].
It is worth noting that this result is in line with previous expectations \citep[e.g.][]{Bisterzo14,Mashonkina17} claiming the [Ba/Eu] as a robust tracer of r-/s-process contribution to chemical enrichment. 

\begin{figure}
    \centering
    \includegraphics[width=0.85\linewidth]{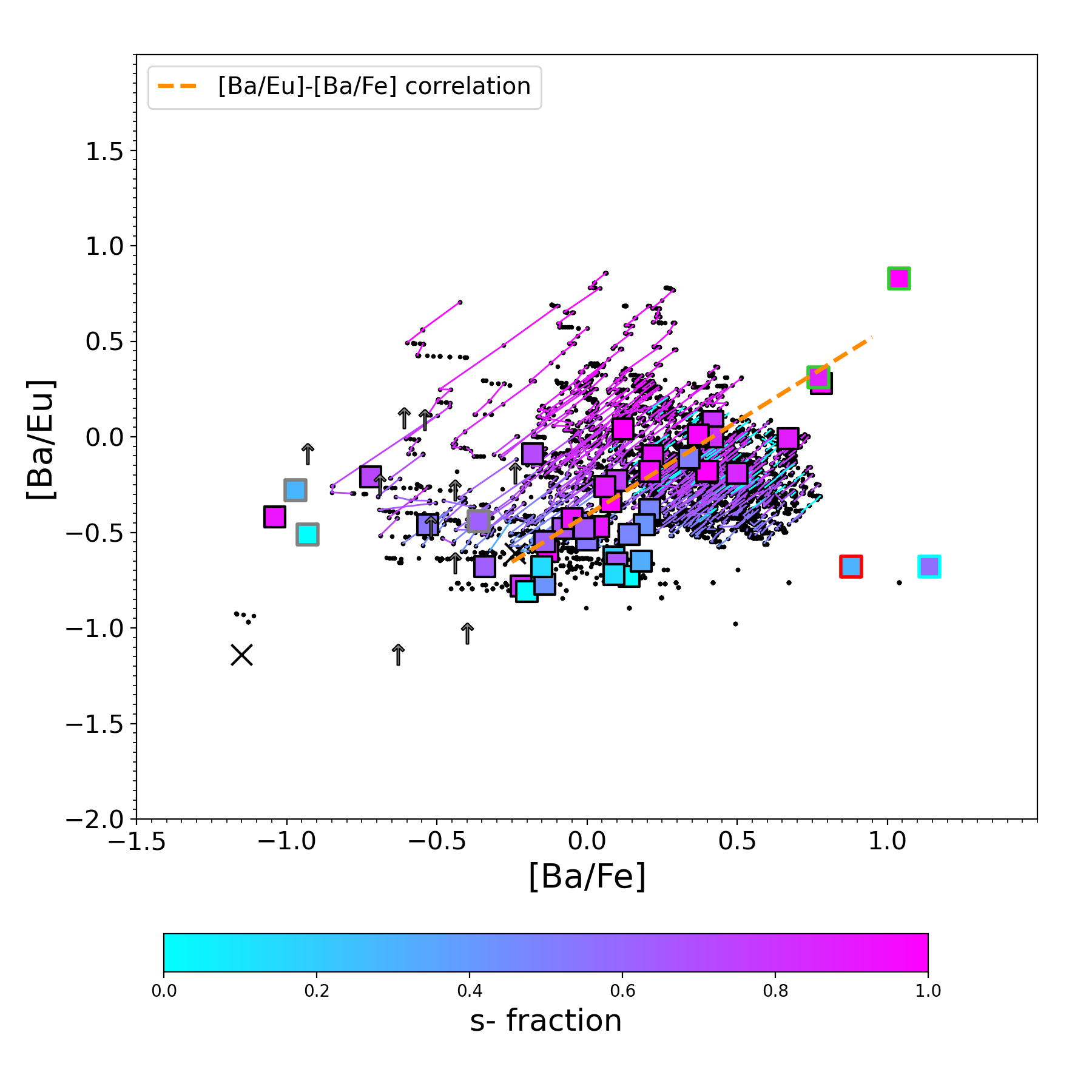}
    \caption{\tiny [Ba/Eu] versus [Ba/Fe] diagram as predicted by 20 randomly selected volumes of our fiducial MW chemical evolution model. Black points are individual timestep predictions by individual volumes, while coloured thin lines are predicted tracks in correspondence of a rising [Ba/Eu] with time.
    Figure colormap displays the fractional s-process contribution to Ba abundance. The dashed orange line reproduce the [Ba/Eu]-[Ba/Fe] data correlation as found in Fig. \ref{fig:BaEu_BaFe}. Data legend is as in Fig. \ref{fig:BaEu_BaFe_sfract}. 
    }
    \label{fig:BaFe_individual_volumes}
\end{figure}

{However, the correlation shown in Fig.~\ref{fig:BaEu_BaFe} exhibits a slope similar to that of chemical evolution tracks of individual stochastic volumes, particularly in the region traced by the observational data at [Ba/Fe] $\gtrsim -0.25$~dex, where [Ba/Eu], [Ba/Fe], and s-fractions increase. This is illustrated in Fig.~\ref{fig:BaFe_individual_volumes}, where we show tracks for 20 randomly selected volumes (coloured thin lines) corresponding to rising [Ba/Eu] trends over time. It is evident that the increase in [Ba/Eu] (and s-fraction) independent of [Ba/Fe] arises from the scatter of individual volume evolutions. However, such individual chemical evolution tracks exhibit clear positive correlations in [Ba/Eu]–[Ba/Fe].
On one hand, the similar slopes observed in Fig.~\ref{fig:BaEu_BaFe} and in the model tracks suggest that the typical [Ba/Eu]–[Ba/Fe] enhancement produced by s-process enrichment events may explain the positive correlation seen in the observed stars. On the other side, the limited size of the observational sample prevents firm conclusions about the large scatter predicted by the models. Additional high-resolution observations with carefully derived isotopic abundances are needed to further clarify the nucleosynthetic processes operating in the early Universe.}

\section{Summary and conclusions}
\label{sec:conclussions}

In this work, we have reassessed the use of barium as a tracer of neutron-capture nucleosynthesis in metal-poor stars, focusing on the consistency between elemental abundance ratios and isotopic diagnostics. While [Ba/Eu] is widely used as a proxy for the relative contributions of the r- and s-processes \citep[see, e.g.,][]{Sneden2008ARA&A..46..241S}, its interpretation at low metallicity is complicated by barium’s dual nucleosynthetic origin and modelling uncertainties affecting strong \ion{Ba}{ii} lines \citep{Travaglio1999ApJ...521..691T, Mashonkina2010A&A...516A..46M}. Barium isotopic ratios, derived from line-profile fitting of resonance lines with hyperfine structure, provide a more direct diagnostic, but their reliability depends critically on the modelling strategy.

We analysed high-resolution, high S/N spectra of metal-poor dwarfs and giants from the \titan\ benchmark samples \citep{giribaldi2021A&A...650A.194G,giribaldi2023A&A...679A.110G}, deriving Ba and Eu abundances and measuring barium isotopic ratios from the \ion{Ba}{ii} 4934~\AA\ resonance line. Both 1D~LTE and 1D~NLTE treatments were explored, revealing systematic biases in purely 1D approaches, especially for giants, where uncertainties in microturbulence, $A(\mathrm{Ba})$, and \teff\ strongly affect line profiles. In particular, 1D~NLTE tends to spuriously favour r-process dominance via A(Ba) underestimations (Sects.~\ref{sec:fiducial} and \ref{sec:NLTE}). We therefore adopt a hybrid strategy: barium abundances are fixed from 1D~LTE subordinate lines, while 1D~NLTE synthesis of the resonance line is used for isotopic-ratio determination. These 1D~LTE abundances rely on accurate atomic data \citep{Gallagher2020A&A...634A..55G,2015PhRvA..91d0501D,2016NatSR...629772D} and recover the solar  chondritic abundance \citep[A(Ba) = $2.18 \pm 0.02$~dex,][]{lodders2009LanB...4B..712L}. In addition, minimal 3D~NLTE corrections are verified for both the Sun and the sample of metal-poor stars ($<0.1$~dex; Fig.~\ref{fig:3DNLTE_cor}).
Our method successfully recovers the isotopic fractions of benchmark peculiar stars dominated by the r-process, namely BPS CS 22892-052 \citep{Sneden94} and BPS CS 31082-001 \citep{hill2002A&A...387..560H}, as well as the mixed r+s process star TYC~6044-714-1 \citep{gull2018ApJ...862..174G,giribaldi2026arXiv260511074G}.

In the [Ba/Eu]–[Ba/Fe] plane, a clear observational pattern emerges, separating r- and s-process-dominated stars (Fig.~\ref{fig:BaEu_BaFe}). The hybrid approach ensures that inferred isotopic fractions consistently trace this sequence for both dwarfs and giants, empirically validating barium isotopic ratios as robust probes of neutron-capture nucleosynthesis. Comparison with stochastic chemical evolution models of the early Milky Way and GES progenitor reproduces both [Ba/Fe] trends and s-process fraction distributions (Fig.~\ref{fig:BaFe_sfract}), highlighting the sensitivity of Ba enrichment to nucleosynthetic yields from massive stars and providing new constraints on the first stellar generations.

Overall, our results demonstrate that, when combined with careful line-formation treatment, barium isotopic ratios provide a powerful complement to elemental abundance ratios in tracing Galactic chemical evolution. Future progress requires larger high-quality stellar samples, which will be essential to exploit this isotopic diagnostic in large spectroscopic surveys and to enable robust quantitative comparisons with theoretical predictions.

\begin{acknowledgements}
The authors thank the anonymous referee for the careful review and constructive comments, which significantly improved the quality and clarity of the manuscript. R.E.G. and  L.M. acknowledge support from INAF through the Large Grants EPOCH and WST, funding for the WEAVE project, the Mini-Grants Checs (1.05.23.04.02), and financial support under the National Recovery and Resilience Plan (PNRR), Mission 4, Component 2, Investment 1.1, Call for tender No. 104 published on 2 February 2022 by the Italian Ministry of University and Research (MUR), funded by the European Union – NextGenerationEU, through the Project ‘Cosmic POT’ (Grant Assignment Decree No. 2022X4TM3H, MUR).
D.V. and S.C. acknowledge funding by the European Union – NextGenerationEU RFF M4C2 1.1 PRIN 2022 project "2022RJLWHN URKA" and by INAF 2023 Theory Grant ObFu 1.05.23.06.06 “Understanding R-process \& Kilonovae Aspects".
Use was made of the Simbad database, operated at the CDS, Strasbourg, France, and of NASA’s Astrophysics Data System Bibliographic Services. 
This publication makes use of data products from the Two Micron All Sky Survey, which is a joint project of the University of Massachusetts and the Infrared Processing and Analysis Center/California Institute of Technology, funded by the National Aeronautics and Space Administration and the National Science Foundation. This research used Astropy,\footnote{http://www.astropy.org} a community-developed core Python package for Astronomy \citep{astropy:2018}.
This work presents results from the European Space Agency (ESA) space mission Gaia. Gaia data are processed by the Gaia Data Processing and Analysis Consortium (DPAC). Funding for the DPAC is provided by national institutions, in particular the institutions participating in the Gaia MultiLateral Agreement (MLA). The Gaia mission website is \url{https://www.cosmos.esa.int/gaia}. The Gaia archive website is \url{https://archives.esac.esa.int/gaia}.
This work is based on observations collected at the European Southern Observatory under ESO programmes:
67.D-0439(A),
095.D-0504(A), 093.D-0095(A), 68.B-0475(A), 074.B-0639(A), 65.L-0507(A),
076.B-0133(A), 077.D-0299(A), 68.D-0094(A), 71.B-0529(A), 67.D-0086(A),
077.B-0507(A), 70.D-0474(A), 072.B-0585(A), 076.A-0463(A), 086.D-0871(A), 090.B-0605(A), 67.D-0554(A), 188.B-3002(H), 188.B-3002(B),
266.D-5655(A), 072.C-0513(B), 075.D-0760(A), 096.C-0053(A), 081.C-0802(C), 082.C-0427(A), 086.C-0145(A), 072.C-0488(E), 073.D-0590(A),
081.D-0531(A), 183.C-0972(A), 183.D-0729(A), 080.D-0347(A), 078.C-0233(B), 082.B-0610(A), 085.C-0063(A), 190.C-0027(A), 192.C-0852(A),
196.C-0042(E), 295.C-5031(A), 60.A-9036(A), 60.A-9700(G), 
086.C-0284(A), 086.C-0448(A), 289.D-5015(A), 
68.D-0546(A), 165.N-0276(A), 0103.A-9013(A), 0104.A-9005(A), 097.A-9024(A), 68.B-0618(A), 65.N-0534(A), 167.D-0173(A), 076.B-0055(A),  083.B-0281(A), 0103.D-0310(A), 0104.D-0059(A), 079.D-0567(A), 078.B-0238(A), 080.D-0333(A),
0103.D-0118(A), 0103.D-0616(A), 165.N-0.276(A), 075.D-0600(A), and 114.27JT.001. 
The 3D NLTE corrections of Ba used in this work were provided by the ChETEC-INFRA project (EU project no. 101008324), task 5.1.
\end{acknowledgements}

\bibliographystyle{aa.bst}

\bibliography{Faint2}

\newpage
\begin{appendix} 

\section{Complementary information to the assessment of the Ba abundance and isotopic ratio fidelity}

\begin{table*}
\caption{Element abundances, barium s-process contributions, and isotopic fractions}
\label{tab:barium}
\centering
\tiny
\begin{threeparttable}
{\fontsize{8pt}{7pt}\selectfont 
\begin{tabular}{lcccccccc}
\hline\hline
Star & [Fe/H] & A(Ba) & Ba s-process contribution & $f_{\rm odd}$ & A(Eu)  & A(Ce)  & A(Sr) & [C/Fe] \\
(SIMBAD id.) & [dex] &[dex]  & [\%] & & [dex] & [dex] & [dex] & [dex]\\
\hline
BD+24 1676 & $-2.24$ & $-0.50 \pm 0.12$ &  --- & --- & $< -1.88$ & $< -0.14$ & $0.52 \pm 0.04$ & 0.12 \\
BD-10 388 & $-2.22$& $-0.28 \pm 0.03$ &  --- & --- & $< -1.75$ & $< -0.09$ & $0.80 \pm 0.05$ & 0.17 \\
BPS CS22166-030 & $-3.11$ & $< -1.31$ &  --- & --- & --- & --- & --- & --- \\
CD-33 3337 & $-1.34$ & $0.94 \pm 0.09$  & $55 \pm 20$ & 0.44 & --- & --- & --- & $-0.02$ \\
CD-35 14849 & $-2.15$ & $-0.51 \pm 0.08$ &  --- & --- & $< -2.26$ & $< 0.33$ & $0.57 \pm 0.06$ & 0.08 \\
HD 16031 & $-1.63$ & $0.52 \pm 0.06$ &  $55 \pm 25$ & 0.44 & $-0.68 \pm 0.07$ & $< 0.07$ & $1.33 \pm 0.06$& 0.16 \\
HD 166913 & $-1.51$ & $0.77 \pm 0.06$  & $76 \pm 24$ & 0.32 & $-0.66 \pm 0.07$ & $< 0.17$ & $1.50 \pm 0.07$& 0.12 \\
HD 218502 & $-1.74$ & $0.22 \pm 0.05$ & $82 \pm _{37}^{18}$ & 0.28 & $-0.66 \pm 0.07$ & $< 0.31$ & $1.16 \pm 0.07$ & 0.08 \\
HD 34328 & $-1.64$ & $0.76 \pm 0.07$  & $93 \pm _{23}^{7}$ & 0.22 & $-0.80 \pm 0.30$ & $< -0.03$ & $1.36 \pm 0.07$ & 0.10 \\
HD 59392 & $-1.53$ & $0.86 \pm 0.07$  & $47 \pm 19$ & 0.48 & $-0.42 \pm 0.13$ & $0.28 \pm 0.23$ & $1.52 \pm 0.07$ & 0.30 \\
BD+02 4651 & $-1.62$ & $0.64 \pm 0.12$  & $100 \pm _{55}^{0}$ & 0.18 & $-0.68 \pm 0.07$ & $< 0.09$ & $1.28 \pm 0.08$ & 0.16 \\
BD+03 740 & $-2.52$ & $-0.97 \pm 0.08$  & --- & --- & $< -1.49$ & $< 0.35$ & $0.12 \pm 0.15$ & $<0.00$ \\
BD-13 3442 & $-2.61$ & $-1.04 \pm 0.08$  & --- & --- & $< -2.80$ & $< -0.59$ & $0.46 \pm 0.08$ & 0.06 \\
BD+26 2621 & $-2.42$ & $-1.17 \pm 0.18$  & --- & --- & $< -2.74$ & $< 0.12$ & $-0.14 \pm 0.07$ & 0.14 \\
BD+26 4251 & $-1.28$ & $1.32 \pm 0.05$  & $80 \pm 11$ & 0.29 & $-0.42 \pm 0.07$ & $0.55 \pm 0.14$ & $1.76 \pm 0.08$& 0.05 \\
BD+29 2091 & $-1.96$ & $0.02 \pm 0.04$  & $0 \pm _{0}^{18}$ & 0.75 &  $-0.83 \pm 0.07$ & $-0.07 \pm 0.10$ & $0.80 \pm 0.07$ & 0.08 \\
CD-30 18140 & $-1.79$ & $0.26 \pm 0.06$  & $100 \pm _{45}^{0}$ & 0.18 & $-0.80 \pm 0.07$ & $0 \pm 0.10$ & $1.10 \pm 0.06$ &--- \\
CD-33 1173 & $-2.72$ & $< -1.61$ &  --- & --- & $< -1.78$ & $< 0.53$ & $-0.30 \pm 0.04$ &--- \\
CD-48 2445 & $-1.88$ & $0.39 \pm 0.04$  & $20 \pm 15$ & 0.64 & $-0.64 \pm 0.07$ & $< 0.07$ & $1.27 \pm 0.08$ & 0.16 \\
CD-71 1234 & $-2.23$ & $-0.74 \pm 0.06$  & --- & --- & $< -2.15$ & $< -0.30$ & $-0.20 \pm 0.05$ & 0.12\\
G 24-3 & $-1.50$ & $0.74 \pm 0.06$ &  $88 \pm _{23}^{12}$ & 0.25 & $-0.66 \pm 0.07$ & $0.27 \pm 0.20$ & $1.30 \pm 0.07$ & $-0.26$ \\
HD 108177 & $-1.58$ & $0.52 \pm 0.07$  & $68 \pm 32$ & 0.36 & $-0.66 \pm 0.18$ & $< 0.19$ & $1.38 \pm 0.06$& $0.15$  \\
HD 116064 & $-1.90$ & $0.14 \pm 0.06$  & $62 \pm _{47}^{38}$ & 0.40 &  $-0.97 \pm 0.49$ & $< 0.01$ & $0.80 \pm 0.07$ & $-0.02$ \\
HD 122196 & $-1.65$ & $0.35 \pm 0.05$  & $71 \pm 24$ &0.34 & $-1.22 \pm 0.60$ & $< 0.06$ & $1.06 \pm 0.06$& $-0.04$ \\
HD 126681 & $-1.22$ & $1.36\pm 0.10$ & $100 \pm _{33}^{0}$ & 0.18 & $-0.12 \pm 0.18$ & $0.71 \pm 0.16$ & $1.76\pm0.06$& 0.98 \\
HD 132475 & $-1.44$ & $1.16 \pm 0.77$  & $82 \pm 16$ & 0.28 & $-0.50 \pm 0.15$ & $0.35 \pm 0.10$ &$1.60 \pm 0.07$ & $0.09$ \\
HD 160617 & $-1.69$ & $0.68 \pm 0.07$  & $42 \pm 19$ & 0.51 & $-0.52 \pm 0.07$ & $0.08 \pm 0.15$ & $1.118 \pm 0.07$ & $0.08$ \\
HD 189558 & $-1.20$ & $1.48 \pm 0.10$  & $84 \pm _{27}^{16}$ & 0.27 & $0 \pm 0.07$ & $0.69 \pm 0.09$ & $1.83 \pm 0.06$ & $-0.10$ \\
HD 193901 & $-1.12$ & $1.20 \pm 0.09$  & $52 \pm 26$ & 0.45 & $0.04 \pm 0.07$ & $0.67 \pm 0.08$ & $1.62 \pm 0.07$ & $-0.08$ \\
HD 213657 & $-1.82$ & $0.31 \pm 0.06$  & $92 \pm _{38}^{8}$ & 0.22 & $-0.92 \pm 0.23$ & $< 0.10$ & $1.00 \pm 0.06$& $0.06$ \\
HD 241253 & $-1.11$ & $1.31 \pm 0.10$  & $84 \pm 27$ & --- & --- & --- & --- \\
HD 284248 & $-1.54$ & $0.50 \pm 0.06$  & $42 \pm 26$ & 0.51 & $-0.39 \pm 0.28$ & $< 0.52$ & $1.26 \pm 0.07$ & 0.00 \\
HD 74000 & $-1.85$ & $0.45 \pm 0.06$  & $100 \pm _{38}^{0}$ & 0.18 & $-1.25 \pm 0.07$ & $< -0.27$ & $1.25 \pm 0.05$& $0.60$ \\
HD 94028 & $-1.36$ & $1.19 \pm 0.04$  & $97 \pm _{9}^{3}$ & 0.19 & $-0.48 \pm 0.07$ & $< 0.57$ & $1.67 \pm 0.08$ & $0.02$ \\
HE 0926-0508 & $-2.50$ & $-0.76 \pm 0.05$  & --- & --- & $< -1.76$ & $< 0.17$ & $0.98 \pm 0.07$ & $0.65$ \\
LP 815-43 & $-2.85$ & --- & --- & --- &  --- & --- & --- & 0.40 \\
LP 831-70 & $-3.14$ & $< -1.72$  & --- & --- & $< -3.00$ & $< 0.38$ & --- & $<0.70$ \\
Ross 453 & $-1.92$ & $-0.26 \pm 0.05$  & ---& --- & $< -1.45$ & $< 0.04$ & $0.68 \pm 0.05$ & 0.11 \\
Ross 892 & $-1.21$ & $1.64 \pm 0.08$  & $89 \pm _{20}^{11}$ & 0.24  & $-0.01 \pm 0.07$ & $0.78 \pm 0.14$ &$1.77 \pm 0.07$ & $-0.26$ \\
UCAC2 20056019 & $-2.64$ & $-0.86 \pm 0.08$  & --- & --- & $< -1.49$ & $< 0.06$ & $0.10 \pm 0.05$& $-0.40$ \\
Wolf 1492 & $-3.04$ & $-1.00 \pm 0.10$  & --- & --- & $< -1.68$ & $< 0.05$ & $-0.21 \pm 0.05$ & --- \\
HD 140283 & $-2.30$ & $-1.27 \pm 0.08$  & --- & --- & $-1.79 \pm 0.10$ & $< -2.00$ & $-0.24 \pm 0.04$& 0.31 \\
HD 84937 & $-2.08$ & $-0.14 \pm 0.05$ & --- & --- & $0.60 \pm 0.07$ & $< -0.40$ & $-0.19 \pm 0.05$& $0.11$ \\
HD 22879 & $-0.86$ & $1.66 \pm 0.06$  & $50 \pm 13$ & 0.46  & $0.10 \pm 0.07$ & $0.69 \pm 0.05$ & $2.04 \pm 0.06$ & $0.06$ \\
HD 298986 & $-1.32$ & $0.86 \pm 0.06$  & $52 \pm 18$ & 0.45 & $-0.26 \pm 0.07$ & $0.39 \pm 0.36$ & $1.40 \pm 0.07$ & $-0.21$ \\
HD 106038 & $-1.36$ & $1.60 \pm 0.08$ & $95 \pm _{19}^{5}$ & 0.21 & $-0.34 \pm 0.07$ & $0.76 \pm 0.10$ & $1.91 \pm 0.07$ & $0.13$ \\
HD 201891 & $-1.10$ & $1.12 \pm 0.08$  & $91 \pm 21$ & 0.23 & $-0.07 \pm 0.07$ & $0.60 \pm 0.52$ & $1.78 \pm 0.08$ & 0.20 \\
HD 102200 & $-1.13$ & $1.19 \pm 0.06$  & $47 \pm 15$ & 0.48 & $0.04 \pm 0.07$ & $0.58 \pm 0.29$ & $1.75 \pm 0.07$& $0.08$ \\
BPS BS 16023-046 & $-2.78$ & $< -1.33$ &  --- & --- & $< -1.2$ & $< 0.24$ & $-0.30 \pm 0.05$& --- \\
BPS BS 16968-061 & $-2.79$ & --- &  --- & --- & $<-1.86$ & $< -0.82$ & $-0.46 \pm 0.18$ &--- \\
BPS CS 22177-009 & $-2.99$ & $< -1.27$  & --- & --- & $< -1.93$ & $< -0.24$ & $-0.52 \pm 0.11$ & --- \\
BPS CS 22953-037 & $-2.83$ & $< -1.34$  & --- & --- & $< -2.30$ & $< 0.04$ & --- & --- \\
BPS CS 29518-043 & $-3.13$ & $< -1.20$  & --- & --- & $< -1.75$ & $< 0.07$ & $-0.14 \pm 0.15$ &--- \\
\hline
BPS BS 17569-049 & $-2.67$ & $-0.39 \pm _{0.21}^{0.12}$  & $67 \pm 8 \pm 10 \pm 12 \pm 11$ & 0.37 & $-1.35 \pm 0.06$ & $-0.96 \pm 0.06$ & $0.14 \pm 0.07$ &$-0.05$ \\
BPS CS 22953-003 & $-2.64$ & $-0.32 \pm _{0.21}^{0.12}$ & $0 \pm _{0}^{15} \pm 15 \pm 12 \pm 11$ & 0.75 & $-1.25 \pm 0.06$ & $-0.75 \pm 0.11$ & $0.30 \pm 0.11$ &0.43 \\
HD 186478 & $-2.31$ & $-0.47 \pm _{0.21}^{0.12}$  & $63 \pm 9 \pm 5 \pm 12 \pm 11$ & 0.39 & $-1.45 \pm 0.08$ & $-0.93 \pm 0.07$ &$0.48 \pm 0.09$ & $-0.41$ \\
BD+09 2870 & $-2.37$ & $-0.91 \pm _{0.21}^{0.12}$  & $73 \pm _{25}^{18} \pm 10 \pm 12 \pm 11$ & 0.33 & $-2.36 \pm 0.12$ & $-1.48 \pm 0.02$ & $0.51 \pm 0.20$ & $-0.38$ \\
BD-18 5550 & $-2.73$ & $-1.39 \pm _{0.21}^{0.12}$  & $10^{50}_{10} \pm 10 \pm 12\pm 11 $ & 0.69 & $-2.41 \pm 0.07$ & $-1.74 \pm 0.16$ & --- & 0.12 \\
BPS CS 22186-025 & $-2.82$ & $-0.79 \pm _{0.21}^{0.12}$  & $14 \pm _{22}^{45} \pm 20 \pm 12 \pm 11$ & 0.67 & $-1.73 \pm 0.06$ & $-1.13 \pm 0.07$ & $0.01 \pm 0.20$ & $-0.50$ \\
BPS CS 22891-209 & $-3.11$ & $-1.46 \pm _{0.21}^{0.12}$  & $38 \pm _{38}^{55} \pm 10 \pm 12 \pm 11$ & 0.53 & $-2.66 \pm 0.06$ & $-1.82 \pm 0.12$ & $-0.04 \pm 0.06$ & $-0.46$ \\
BPS CS 22896-154 & $-2.64$ & $-0.37 \pm _{0.21}^{0.12}$ & $13 \pm _{13}^{44} \pm 9 \pm 12 \pm 11$ & 0.68 & $-1.31 \pm 0.06$ & $-0.83 \pm 0.10$ & $0.22 \pm 0.09$& 0.43 \\
BPS CS 29491-053 & $-2.81$ & $-1.49 \pm _{0.21}^{0.12}$  & $29 \pm _{29}^{62} \pm 30 \pm 12 \pm 11$ & 0.59 &  $-2.72 \pm 0.09$ & $-2.03 \pm 0.16$ &$-0.12 \pm 0.07$ & $-0.37$ \\
BPS CS 29518-051 & $-2.52$ & $-0.70 \pm _{0.21}^{0.12}$  & $62 \pm _{62}^{31} \pm 10 \pm 12 \pm 11$ & 0.40 & $-1.92 \pm 0.09$ & $-1.00 \pm 0.09$ & $0.29 \pm 0.07$ & $-0.01$ \\
HD 122563 & $-2.73$ & $-1.59 \pm _{0.21}^{0.12}$  & $92 \pm _{39}^{8} \pm 8 \pm 12 \pm 9$ & 0.22 &  $-2.83 \pm 0.11$ & --- &$-0.12 \pm 0.08$ & $-0.50$ \\
CD-41 15048  & $-1.99$ & $0.40 \pm _{0.21}^{0.12}$  & $94 \pm 4 \pm 4 \pm 12 \pm 11$ & 0.21 & $-1.08 \pm 0.07$ & $-0.36 \pm 0.06$ & $0.52 \pm 0.06$ &$-0.32$ \\
BD+17 3248  & $-1.92$ & $0.44 \pm _{0.21}^{0.12}$  & $32 \pm _{4}^{1} \pm 6 \pm 12 \pm 11$ & 0.57 & $-0.57 \pm 0.06$ & $-0.15 \pm 0.06$ & $0.75 \pm 0.08$ &$-0.31$ \\
HD 218857  & $-1.90$ & $-0.31 \pm _{0.21}^{0.12}$  & $65 \pm _{31}^{13} \pm 5 \pm 12 \pm 11$ & 0.38 &  --- & --- & $0.75 \pm 0.09$& --- \\
HD 45282  & $-1.43$ & $0.73 \pm _{0.21}^{0.12}$  & $65 \pm _{31}^{3} \pm 4 \pm 12 \pm 11$ & 0.38 & $-0.46 \pm 0.08$ & $0.19 \pm 0.06$ & $1.35 \pm 0.08$  &$-0.14$ \\
\hline
BPS CS 22892-052 $\dagger$ (CEMP-r)  & $-2.86$ & $0.20 \pm _{0.21}^{0.12}$ & $31 \pm _{1}^{2} \pm 31 \pm 12 \pm 11$ & 0.57 &  $-0.78 \pm 0.06$  & $-0.40 \pm 0.06$ & $0.47 \pm 0.06$ & 1.07 \\
BPS CS 31082-001 $\spadesuit$ (r)  & $-2.61$ & $0.66 \pm _{0.21}^{0.12}$ & $36 \pm _{3}^{2} \pm 3 \pm 12 \pm 11$ & 0.42 & $-0.27 \pm 0.06$ & $-0.09 \pm 0.07$ &$0.51 \pm 0.06$ & 0.27 \\
HD 196944 (CEMP-r+s) & $-2.16$ & $1.06 \pm _{0.21}^{0.12}$  & $100 \pm 0 \pm 5 \pm 12 \pm 11$ &0.18 &  $-1.43 \pm 0.06$ & $0.42 \pm 0.07$  & $1.28 \pm 0.08$& 1.45 \\
TYC 6044-714-1 (CEMP-r+s)  & $-2.17$ & $0.78 \pm _{0.04}^{0.04}$  & $86 \pm 1 \pm 2 \pm 12 \pm 11$ & 0.26 & $-1.19 \pm 0.08$ & $0.20 \pm 0.08$ & $1.00 \pm 0.09$ & 1.00 \\
\hline
\end{tabular}
}
\begin{tablenotes}
\item{} \textbf{Notes.} {The first part of the table lists the \titan~I dwarfs, the second part lists the \titan~II giants,  and the third part includes peculiar stars. 
The symbol ($\spadesuit$) indicates the Hill star. The symbol ($\dagger$) indicates the Sneden star. 
Categories of the peculiar stars according to the literature (see main text in Sect.~\ref{sec:data}) are indicated aside their catalogue numbers.
For giants, s-process contribution uncertainties from different sources are reported separately: from the $v_{mic}$ uncertainty, spectral-noise–induced flux fluctuations, the \teff\ uncertainty,  and $\sigma(\rm{A(Ba)}) = 0.04$~dex (internal error); following this order.
} 
\end{tablenotes}
\end{threeparttable}
\end{table*}

\subsection{Atmospheric parameters}

In Table~\ref{tab:parameters} we list the atmospheric parameters for the stellar sample presented in Table~\ref{tab:barium}. 
Together with the quantities \teff, \logg, [Fe/H], and $v_{mic}$, we present the signal-to-noise (S/N) ratio obtained for each stellar spectra and the high-resolution spectrograph adopted for the observations among UVES, HARPS, and ESPRESSO.

\begin{table*}[h]
\caption{Stellar parameters of the stars}
\label{tab:parameters}
\centering
\tiny 
\begin{threeparttable}
{\fontsize{8pt}{9pt}\selectfont 
\begin{tabular}{lcccccccccc}
\hline\hline
Star & \teff\ &  
\logg & [Fe/H]$_{NLTE}$ & [Fe/H]$_{LTE}$ & $v_{mic}$ & S/N & Spectrum & Pop.\\
(SIMBAD identifier) & [K] & [dex] & [dex] & [dex] & [km/s] \\
\hline
CD-33 3337 & $6044 \pm 43$ & $3.97 \pm 0.04$ & $-1.34 \pm 0.06$ & --- & 1.37 & 462 & UV & discs\\
HD 16031 & $6276 \pm 10$ & $4.27 \pm 0.04$ & $-1.63 \pm 0.05$ & --- & 1.35 & 424 & UV & discs\\
\vdots & \vdots & \vdots & \vdots & \vdots & \vdots & \vdots & \vdots\\
\hline
BD-18 5550 & $4981 \pm 85$ & $1.64 \pm 0.16$ & $-2.73 \pm 0.10$ & $-2.95$ & 1.42 & 520  & UV & proto-MW\\
HD 122563 & $4627 \pm 13$ & $1.06 \pm 0.16$ & $-2.73 \pm 0.09$ & $-2.96$ & 2.46 & 318 & UV & proto-MW\\
\vdots & \vdots & \vdots & \vdots & \vdots & \vdots & \vdots & \vdots\\
\hline
BPS CS 22892-052 $\dagger$ (CEMP-r) & $4979 \pm 38$ & $1.68 \pm 0.16$ & $-2.86 \pm 0.08$ & $-2.96$ & 1.70 & 142 & UV & proto-MW\\
BPS CS 31082-001 $\spadesuit$ (r) & $5028 \pm 8$ & $2.26 \pm 0.16$ & $-2.61 \pm 0.10$  & $-2.77$  & 1.57 & 412  & UV & proto-MW\\
HD 196944 (CEMP-rs) & $5539 \pm 41$  & $2.27 \pm 0.16$  & $-2.16 \pm 0.07$  & $-2.26$  & 2.23 & 232 & HA & proto-MW\\
TYC 6044-714-1 (CEMP-r+s) & $4810 \pm 41$ & $1.70 \pm 0.16$  & $-2.17 \pm 0.05$  &  $-2.27$ & 1.82 & 410 & UV & proto-MW\\
\hline
\end{tabular}
}
\begin{tablenotes}
\item{} \textbf{Notes.} {Stars are listed as in Table~\ref{tab:barium}. 
In the column number eight, UV means UVES, HA HARPS, and ES ESPRESSO. 
The last column, indicates the population to which the star belongs.
The symbol ($\spadesuit$) indicates the Hill star. The symbol ($\dagger$) indicates the Sneden star.   Categories of the peculiar stars according to the literature (see main text in Sect.~\ref{sec:data}) are indicated aside their catalogue numbers.
The table is accessible in a machine-readable format at the CDS. A portion is shown to illustrate its format and content.
} 
\end{tablenotes}
\end{threeparttable}
\end{table*}

\begin{table}
\caption{1D~NLTE barium abundance}
\label{tab:barium_NLTE}
\centering
\tiny
\begin{threeparttable}
\begin{tabular}{lccc}
\hline\hline
Star &  A(Ba)$_{NLTE}$  \\
(SIMBAD id.) & [dex] \\
\hline
CD-33 3337 &  0.78  \\
HD 16031 &   0.48  \\
\vdots & \dots \\
\hline
BD-18 5550 &  $-1.45$  \\
HD 122563 &  $-1.56$  \\
\vdots & \dots \\
\hline
BPS CS 22892-052 $\dagger$ (CEMP-r)  & $-0.08$  \\
BPS CS 31082-001 $\spadesuit$ (r)  & 0.30  \\
HD 196944 (CEMP-rs) &  0.60  \\
TYC 6044-714-1 (CEMP-r+s) &  0.46  \\
\hline
\end{tabular}
\begin{tablenotes}
\item{} \textbf{Notes.} {Symbols are as in Table~\ref{tab:barium}.
The table is accessible in a machine-readable format at the CDS. A portion is shown to illustrate its format and content.
} 
\end{tablenotes}
\end{threeparttable}
\end{table}

\begin{table}
\centering
\tiny
\caption{Ba resonance subordinate lines at 6141 and 6496~\AA\ with isotopic and hyperfine structure splitting}
\begin{tabular}{cccccc}
\hline\hline
Isotope & Wavelength [\AA] & log $gf$\\
\hline
$^{134}$Ba & 6141.7295  & $-0.032$ \\
\\
\multirow{9}{*}{$^{135}$Ba} & 6141.7231 & $-2.419$ \\
& 6141.7246 & $-1.264$\\
& 6141.7261 & $-2.226$\\
& 6141.7271 & $-0.459$\\
& 6141.7271 & $-1.170$\\
& 6141.7290 & $-0.664$\\
& 6141.7290 & $-1.282$\\
& 6141.7300 & $-0.913$\\
& 6141.7305 & $-1.235$\\
\\
$^{136}$Ba & 6141.7280 & $-0.032$\\
\\
\multirow{9}{*}{$^{137}$Ba}  &6141.7183 & $-2.419$\\
&6141.7202 & $-1.264$\\
&6141.7222 & $-2.226$\\
&6141.7231 & $-0.496$\\
&6141.7231 & $-1.170$\\
&6141.7251 & $-0.664$\\
&6141.7251 & $-1.282$\\
&6141.7266 & $-0.913$\\
&6141.7266 & $-1.235$\\
\\
$^{138}$Ba & 6141.7271 & $-0.032$\\
\hline
$^{134}$Ba & 6496.9102 & $-0.407$\\
\\
\multirow{6}{*}{$^{135}$Ba}  & 6496.8989 & $-1.913$\\
& 6496.9019 & $-1.212$ \\
&6496.9067 & $-0.765$ \\
&6496.9160 & $-1.610$ \\
&6496.9175 & $-1.212$ \\
&6496.9204 & $-1.212$ \\
\\
$^{136}$Ba & 6496.9102 & $-0.407$\\
\\
\multirow{6}{*}{$^{137}$Ba}  & 6496.8979 & $-1.913$\\
&6496.9014 & $-1.212$\\
&6496.9062 & $-0.765$\\
&6496.9165 & $-1.610$\\
&6496.9185 & $-1.212$\\
&6496.9219 & $-1.212$\\
\\
$^{138}$Ba & 6496.9102& $-0.407$\\
\hline
\end{tabular}
\label{tab:tab1}
\end{table}

\subsection{1D~LTE versus 1D~NLTE in Ba abundance and isotopic ratio determination}
\label{sec:NLTE}

\begin{figure*}
    \centering
    \includegraphics[width=0.8\linewidth]{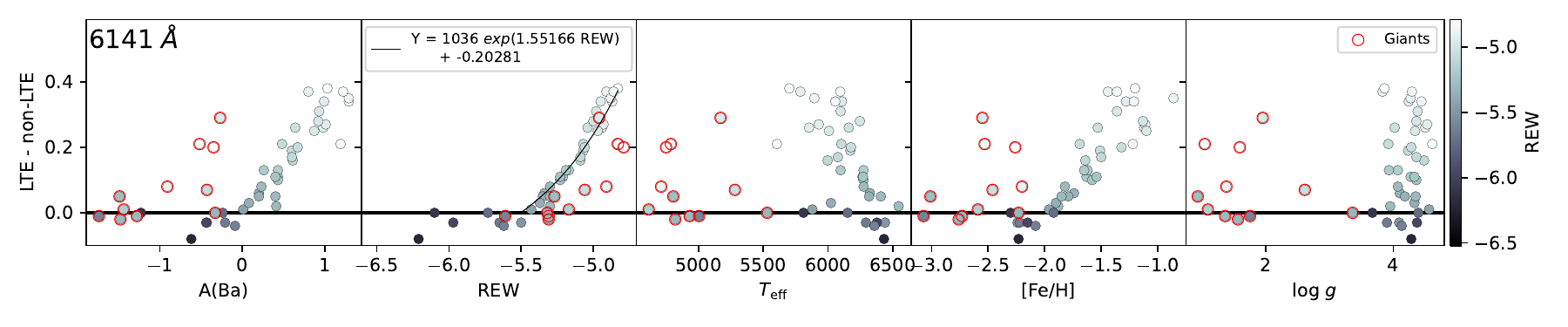}
    \includegraphics[width=0.8\linewidth]{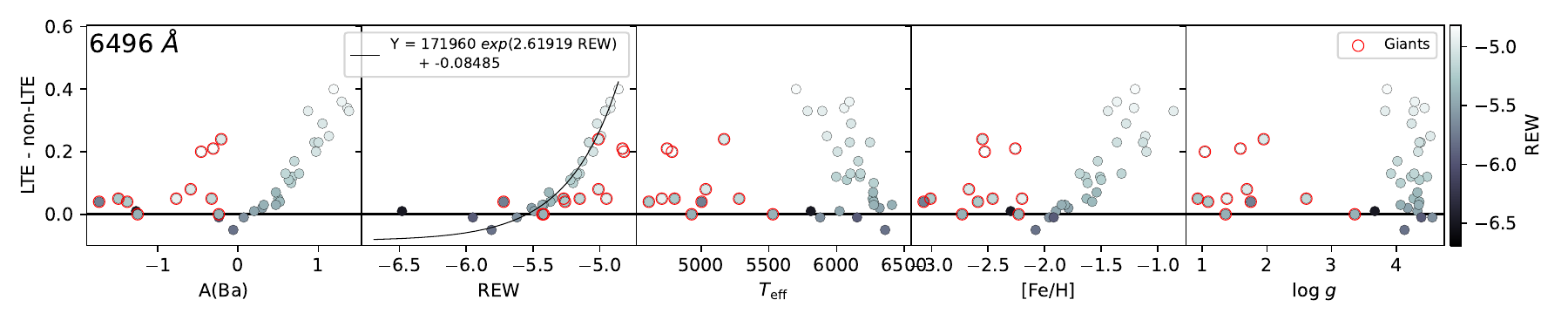}
    \caption{\tiny Barium abundance differences between 1D~LTE and 1D~NLTE synthesis for the lines 6141 and 6496~\AA. The abundance differences are shown as a function of the atmospheric parameters and reduced equivalent width (REW). Symbols are colour-coded according to REW. Giant stars are highlighted with red contours.}
    \label{fig:Ba_NLTE}
\end{figure*}

We present 1D~NLTE Ba abundances for the stars of the adopted sample in Table~\ref{tab:barium_NLTE} to allow a direct confrontation with 1D~LTE abundances in Table~\ref{tab:barium}.
Figure.~\ref{fig:Ba_NLTE} shows comparisons for the lines 6141 and 6496~\AA.  Well defined exponential trends appear as function of the reduced equivalent width (REW = log($\rm{EW}/\lambda$)\footnote{EW is the equivalent width in m\AA\ and $\lambda$ is the wavelength in \AA.}) leading to differences of up to 0.4~dex for the strongest lines. REW is correlated with A(Ba), and with [Fe/H] due to Galaxy evolutionary effects; therefore correlations of the differences with these parameters are evident.

\begin{figure*}
    \centering
    \includegraphics[width=0.8\linewidth]{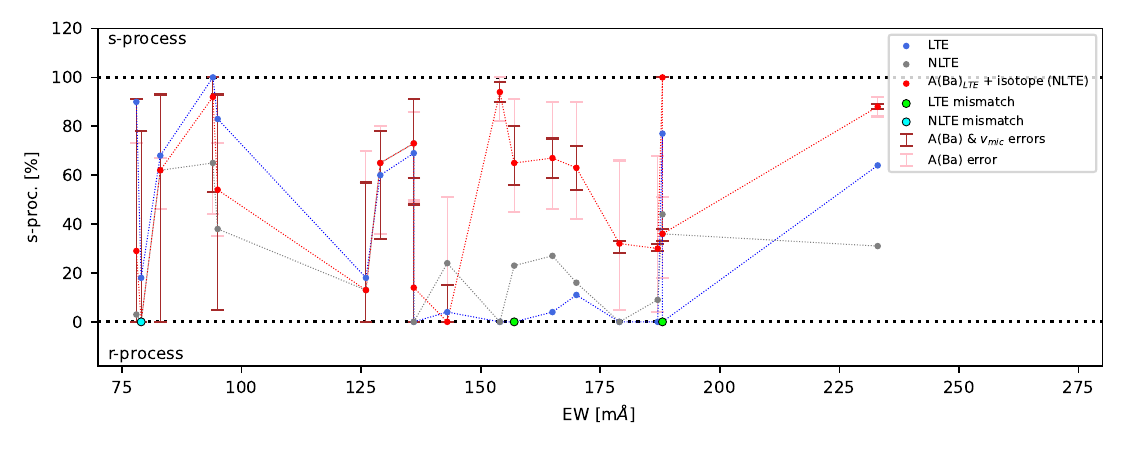}
    \caption{\tiny Isotopic ratios expressed as s-process fractions as a function of equivalent width. Three sets of determinations are shown: both A(Ba) and the isotopic ratios derived under 1D~LTE (blue points), both derived under 1D~NLTE (gray points), and A(Ba) derived under 1D~LTE with isotopic ratios derived under 1D~NLTE (red points). Error bars are displayed for the latter case: pink bars are related to typical uncertainties in A(Ba), while brown bars show the combined effects of uncertainties in $v_{\mathrm{mic}}$ and A(Ba). Green circles indicate 1D~LTE fits whose synthetic profiles are significantly shallower than the observations, thus not adjustable. Thin dotted lines connect points of the same colour to guide the eye.
    }
    \label{fig:EW_sproc}
\end{figure*}

Figure~\ref{fig:EW_sproc} shows the s-process fractions in giant stars as a function of EW. 
Three sets of determinations are compared: 
The first is obtained under 1D~LTE using 1D~LTE-derived A(Ba) values (“full LTE”, blue points).
The second is obtained under 1D~NLTE using 1D~NLTE-derived A(Ba) values (“full NLTE”, grey points).
The third is obtained under 1D~NLTE while fixing A(Ba) to the 1D~LTE value (“fiducial diagnostic”, red points), which we adopt as our reference throughout this work.
Focusing on the stars with the most precise isotopic determinations (EW~$\gtrsim 140$~m\AA), we observe that the fiducial diagnostic differs substantially from both the full LTE and full NLTE ones, leading to opposite conclusions regarding the dominant nucleosynthetic process (r or s). In some cases, full LTE modelling is clearly inadequate: the two green points in the figure correspond to the Hill star and HD~45282, for which 1D~LTE synthetic line profiles are far too shallow to reproduce the observations.
When this happens, determination methods (e.g. EW comparison or line profile fitting) bias to r-process diagnoses.
The error bars in the figure show that for giants with EW $\gtrsim 140$~m\AA, the uncertainties are lower than $10\%$, whereas for EW $\lesssim 140$~m\AA, they mostly exceed 50\%, despite the high precision of their atmospheric parameters and high S/N spectra.
We therefore recommend excluding giants with lines in this EW regime from isotopic analyses.
Notably, the combined uncertainties from $v_{mic}$ and A(Ba) variations are smaller than those obtained considering only the internal A(Ba) errors, because the opposing effects partially cancel.

\begin{figure*}
    \centering
    \includegraphics[width=0.32\linewidth]{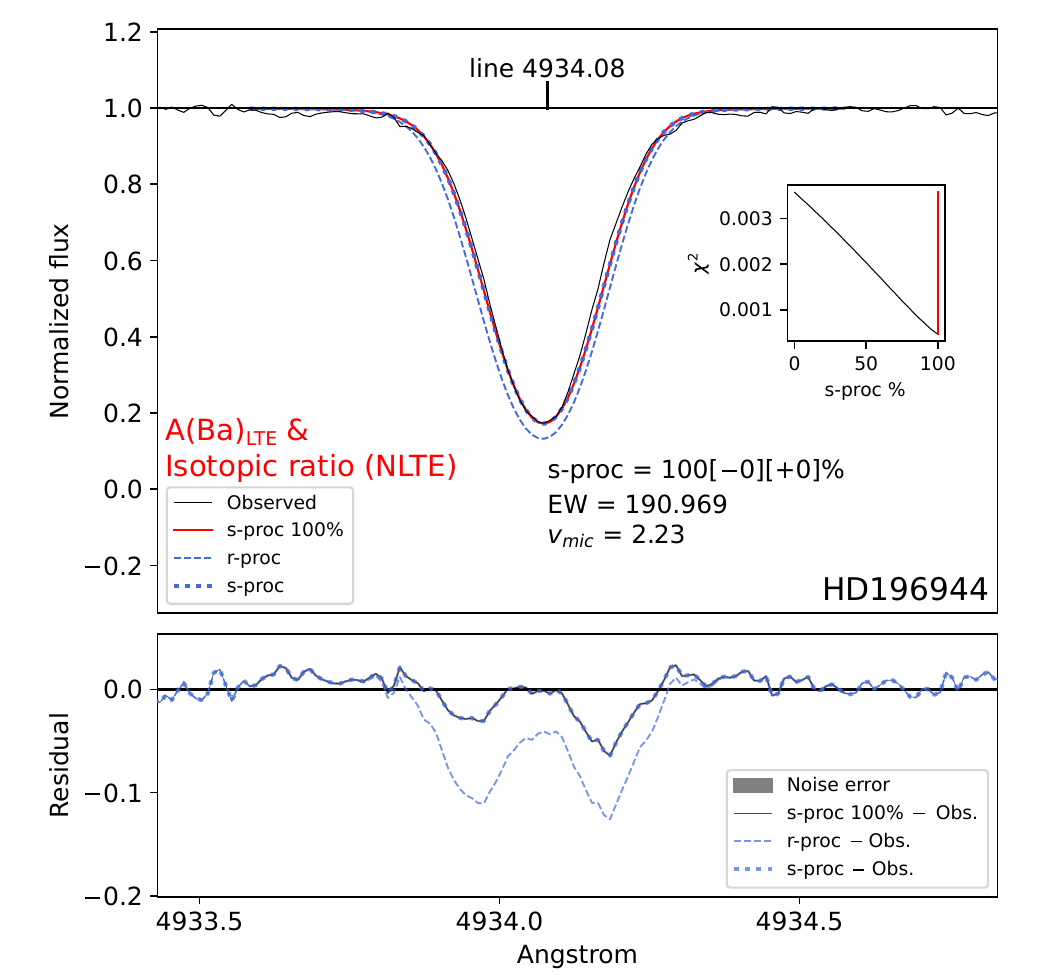}
    \includegraphics[width=0.32\linewidth]{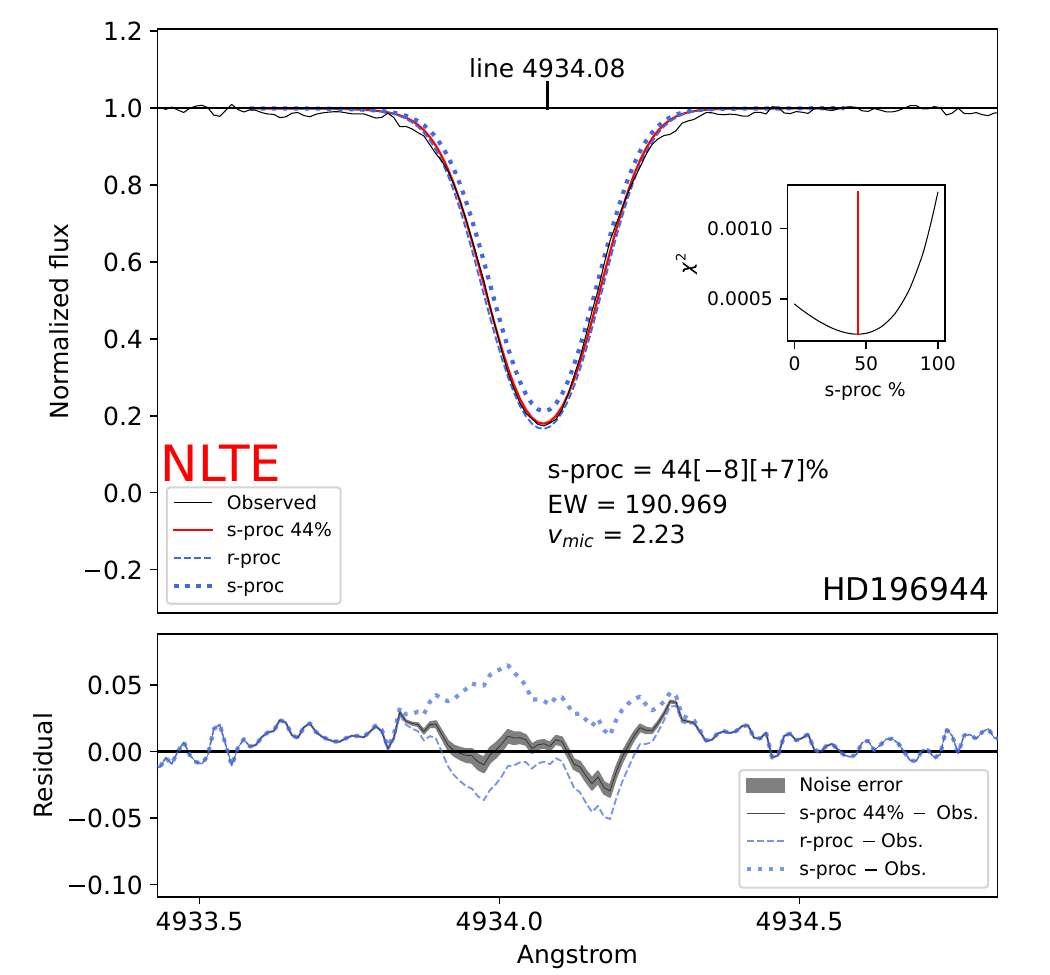}
    \includegraphics[width=0.32\linewidth]{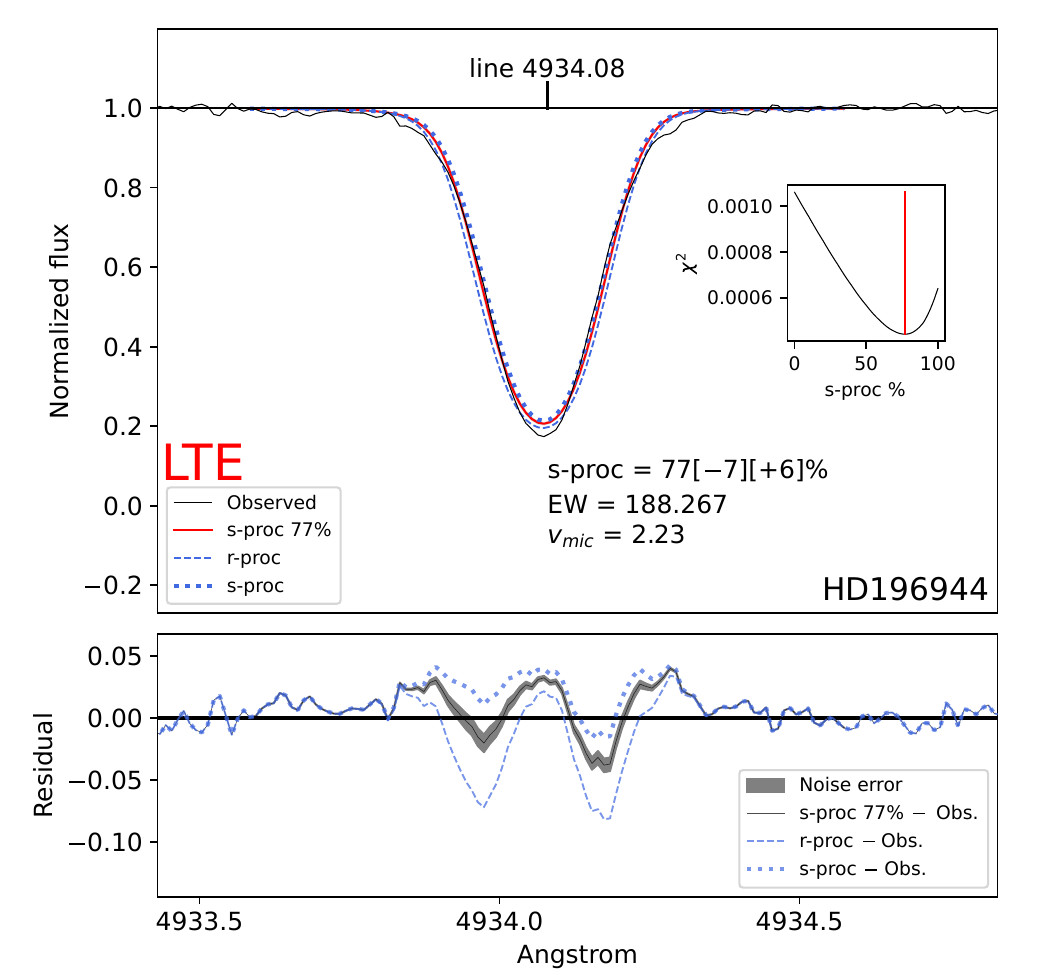}\\
    \includegraphics[width=0.32\linewidth]{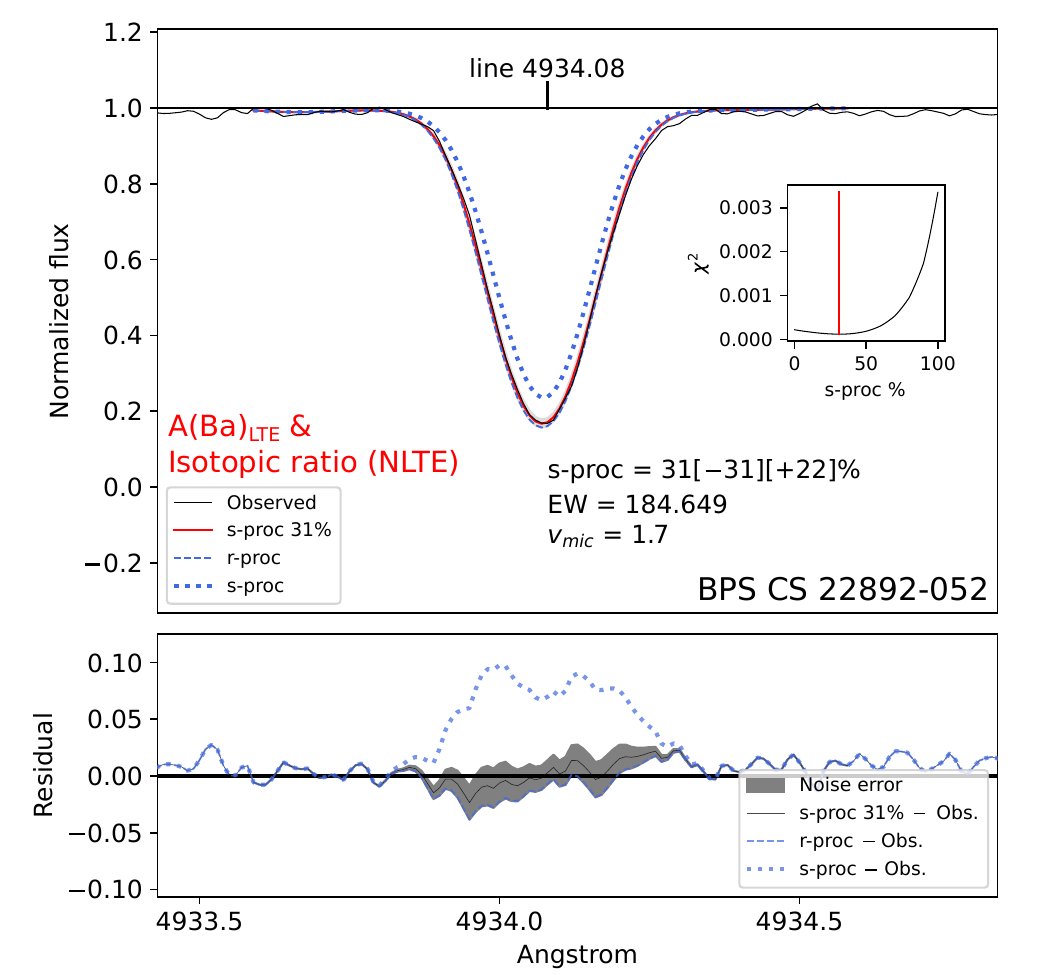}
    \includegraphics[width=0.32\linewidth]{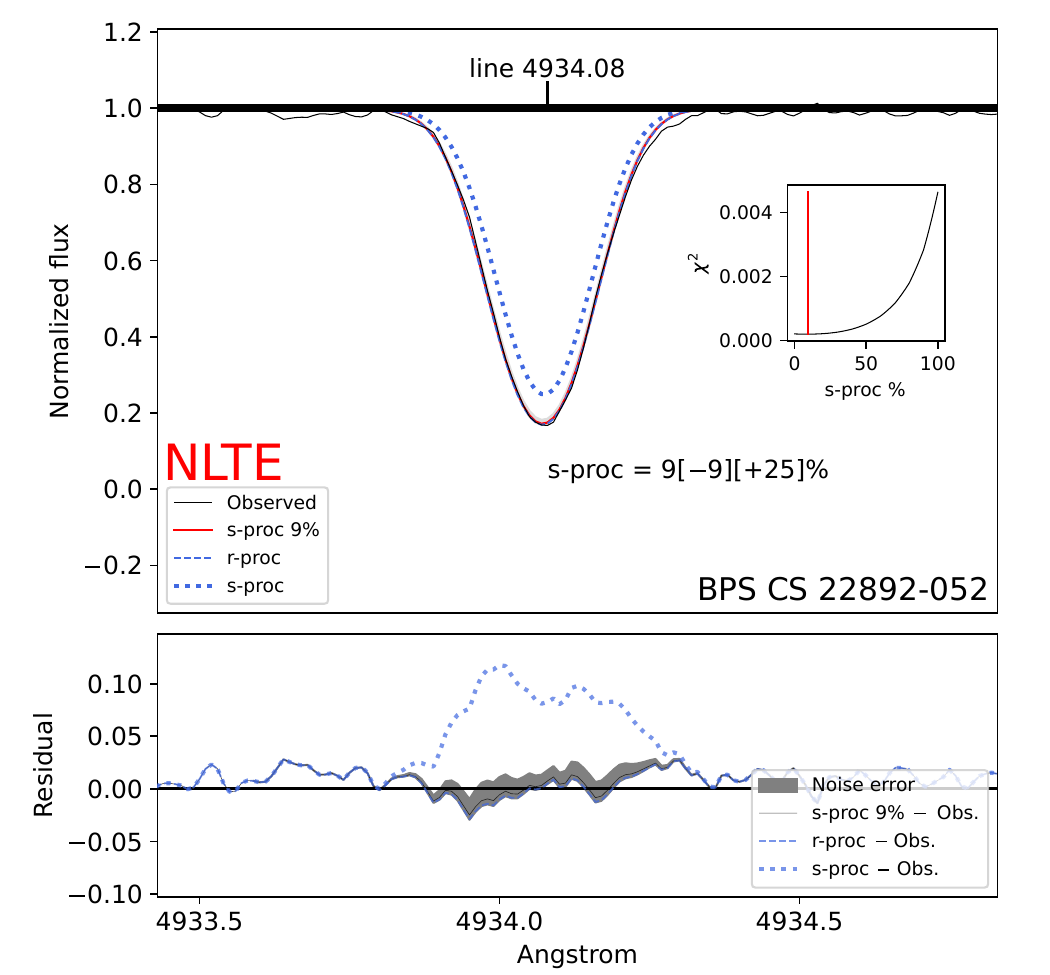}
    \includegraphics[width=0.32\linewidth]{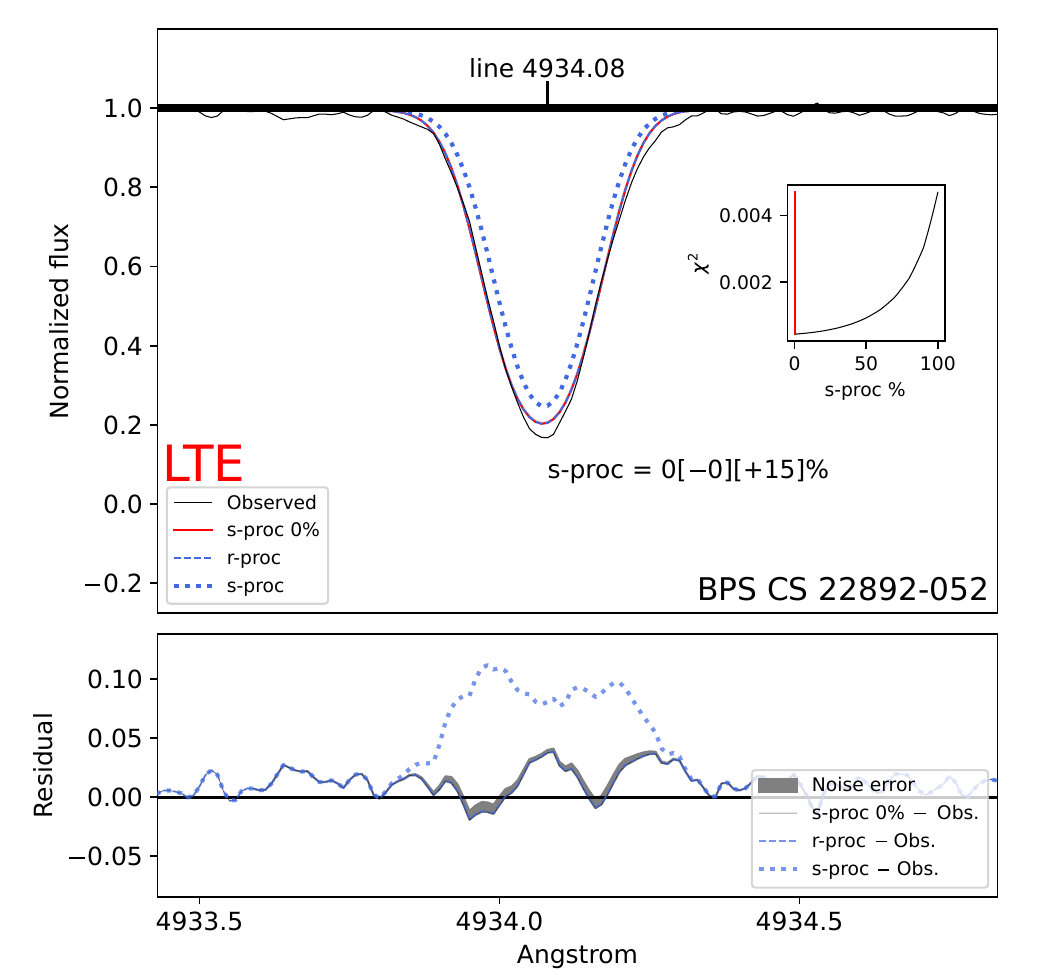}
    \caption{\tiny { Fits to the 4934~\AA\ line profile of HD~196944 {\it (top panels)} and the Sneden star {\it (bottom panels)}. From the left to the right, results from the fiducial diagnostic, full 1D NLTE and full 1D LTE approaches are displayed. 
    Synthetic profiles corresponding to pure s- and r-processes are shown as indicated in the legends.
    The best-fit synthetic profile is plotted in red, and its derived s-process fraction and the associated noise-related uncertainty are reported. The inset displays the $\chi^{2}$ distribution used to determine the optimal fit, with the minimum marked by a vertical red line.
    As in Fig.~\ref{fig:HD160617}, the bottom panels show the residuals of the best-fit profile, the pure-s model, and the pure-r model relative to the observed profile.}
    }
    \label{fig:giant_fits}
\end{figure*}

This experiment demonstrates that both full 1D LTE and full 1D NLTE analyses bias isotopic-ratio diagnostics in giant stars toward an r-process signature. In the 1D LTE case, the synthetic resonance line is too shallow, artificially increasing the contribution of the odd isotopes and simulating certain r-process influence. 
Such effect is exemplified in Fig.~\ref{fig:giant_fits} with the profile of HD~196944, dominated by the s-process, and with the Sneden star,  dominated by the r-process \citep{Sneden94}: compare fits in the right and left panels. 
Although for the Sneden star none of the three syntheses mistake the r-process dominance, it is clear that the full 1D~LTE approach is not able to properly model the line core.
In the full 1D NLTE case, the inferred A(Ba) is significantly underestimated, which similarly shifts the solution toward an r-process-like composition.
This is more clearly shown for HD~196944 in the middle top panel.

\subsection{The case of the CEMP star HD~196944}
\label{sec:hd196944}

HD~196944 was classified as an r/s-process star by \citet[][hereafter K21]{karinkuzhi2021A&A...645A..61K}  and as a likely r- or i-process star by \citet[][hereafter S25]{sitnova2025A&A...704A.103S}. In contrast, our analysis  indicates that the s-process is the dominant contributor to its barium enrichment, see top left panel of Fig.~\ref{fig:giant_fits}. We examine the methodological differences among the studies.

\begin{figure}
    \centering
    \includegraphics[width=0.8\linewidth]{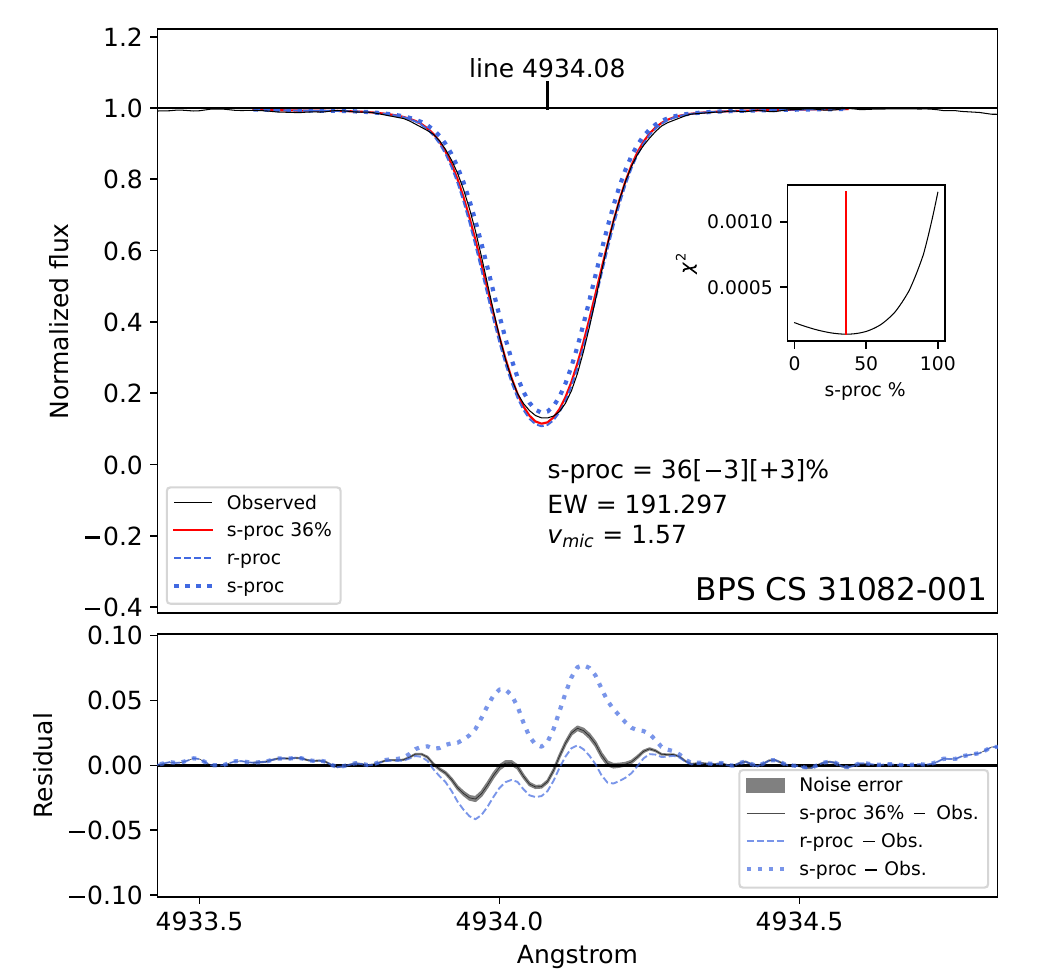}
    \caption{\tiny Similar to left panels in Fig.~\ref{fig:giant_fits}, but for the Hill star.
    }
    \label{fig:peculiars}
\end{figure}

\begin{table}[!htbp]
\caption{Atmospheric parameters adopted for HD~196944 in different studies}
\tiny
\label{tab:HD196944}
\centering
\begin{threeparttable}
\begin{tabular}{l|ccccccccc}
\hline\hline
Authors & \teff\ & \logg\ & [Fe/H] & $v_{mic}$ \\
\hline
K21  & $5168 \pm 48$ & $1.28$ &  $-2.50$  & $1.68$  \\
S25 & $5390 \pm 80$ & $2.13$ & $-2.21$ & $2.00$\\
G23 & $5539 \pm 41$ & 2.44 & $-2.05$ & -- \\
This work & $5539 \pm 41$ & $2.27$ & $-2.26$ & 2.23\\
\hline
\end{tabular}
\begin{tablenotes}
\item{} \textbf{Notes.} {K21: \cite{karinkuzhi2021A&A...645A..61K}, S25: \cite{sitnova2025A&A...704A.103S}, G23: \cite{giribaldi2023A&A...679A.110G}. 
} 
\end{tablenotes}
\end{threeparttable}
\end{table}

K21 base their classification on comparisons between observed heavy-element abundances ($Z>56$) and nucleosynthesis predictions. In their Fig.~11, the $Z$–[X/Fe] distribution shows a mildly negative slope for $56 \leq Z \leq 70$, which they interpret as possible indicative of mixed r+s contributions, whereas steeper slopes are associated with s-process dominance. They report [Ba/Eu] = 0.24~dex, while our value is higher by approximately~0.5 dex, which, within the same diagnostic framework, favours a stronger s-process contribution.
Part of the discrepancy may be related to the adopted atmospheric parameters. As shown in Table~\ref{tab:HD196944}, the parameters used by K21 display large offsets expected when excitation and ionisation equilibrium of Fe lines is imposed under 1D~LTE in giant stars \citep{ruchti2013MNRAS.429..126R, giribaldi2023A&A...679A.110G}.
Tracing the exact impact on the derived abundances would require a line-by-line reassessment, since parameter offsets affect different transitions unevenly. For instance,  \cite{giribaldi2026arXiv260511074G} shows that, in metal-poor red giants, typical uncertainties of 50~K in \teff\ and 0.15~dex in \logg\ can induce relative abundance shifts of 0.01–0.04 dex between neighbouring elements (e.g. Ba–La–Ce or Al–Mg–Si).

S25 adopt a complementary approach, relating the odd-isotope fraction of Ba to [Ba/Eu]. 
They adopt the log~$gf$ values of \cite{miles1969AD......1....1M}, which when applied to the 6141 and 6496~\AA\ Ba lines in the solar spectrum, yield A(Ba) determinations higher than the chondritic (A(Ba)$_\odot = 2.18 \pm 0.02$~dex) by +0.5 dex in 1D~LTE and +0.3 dex in 1D~NLTE.
This explains why they obtain A(Ba) values higher than ours by +0.35 under 1D~LTE, but  nearly equal to ours under 1D~NLTE, from atmospheric parameters similar to those here (see Table~\ref{tab:HD196944}).
They obtain an odd-isotope fraction ($f_{\rm odd} \approx 0.62$) that favours a stronger r-process contribution, or alternatively the i-process.
Since both our and their analyses rely on the same observed spectrum, the differing results likely arise from a combination of factors: 1D~NLTE calculations from different sources, different radiative transfer codes, adopted atomic data, and isotopic-ratio derivation methods. 
We suspect that the latter may be relevant. Namely, theirs  consists on matching EW of synthetic and observational spectra, whereas we perform line fitting; the results of which may diverge for over-saturated lines.   
The radiative transfer treatment and the adopted log~$gf$ values for the 4934~\AA\ resonance line and its HFS components also differ between the studies. S25 use $-0.150$ \citep{miles1969AD......1....1M}, while we adopt $-0.172$ \citep[Table B.1. in ][and references in the paper]{giribaldi2025A&A...702A..65G}.

\subsection{The cases of the Sneden and Hill stars}
\label{sec:peculiars}

The Sneden (BPS~CS~22892-052) and Hill (BPS~CS~31082-001) stars are classic r-process-dominated stars, with [Eu/Fe] = 1.56 and 1.82~dex, respectively.
Bottom left panel in Fig.~\ref{fig:giant_fits} and Fig.~\ref{fig:peculiars} show the profile fits from our fiducial diagnostic. 
Both cases display over-saturated lines with EW $\sim190$, where the inset $\chi^2$ subplots show that the likelyhood variation from 0\% to $\sim$60\% s-process contribution is minimal. 
The errors in Table~\ref{tab:barium}  conciliate  $\approx$0\% s-process contributions. However, it is possible that such small systematics (to higher s-process contributions) are related to subtle granulation effects lacking in 1D~NLTE modelling \citep[e.g.][]{frame2025MNRAS.539.2248F}. 
Anyhow, we remark that, for s-process dominated stars with over-saturated lines, the s-process percentage contribution can be accounted with  good precision, as in the cases of HD~196944  (Sect.~\ref{sec:hd196944}) and TYC~6044-714-1  \citep{giribaldi2026arXiv260511074G}, for example.

\subsection{The case of BD$-18\,5550$}
\label{sec:BD18}
\begin{figure*}
    \centering
    \includegraphics[width=0.4\linewidth]{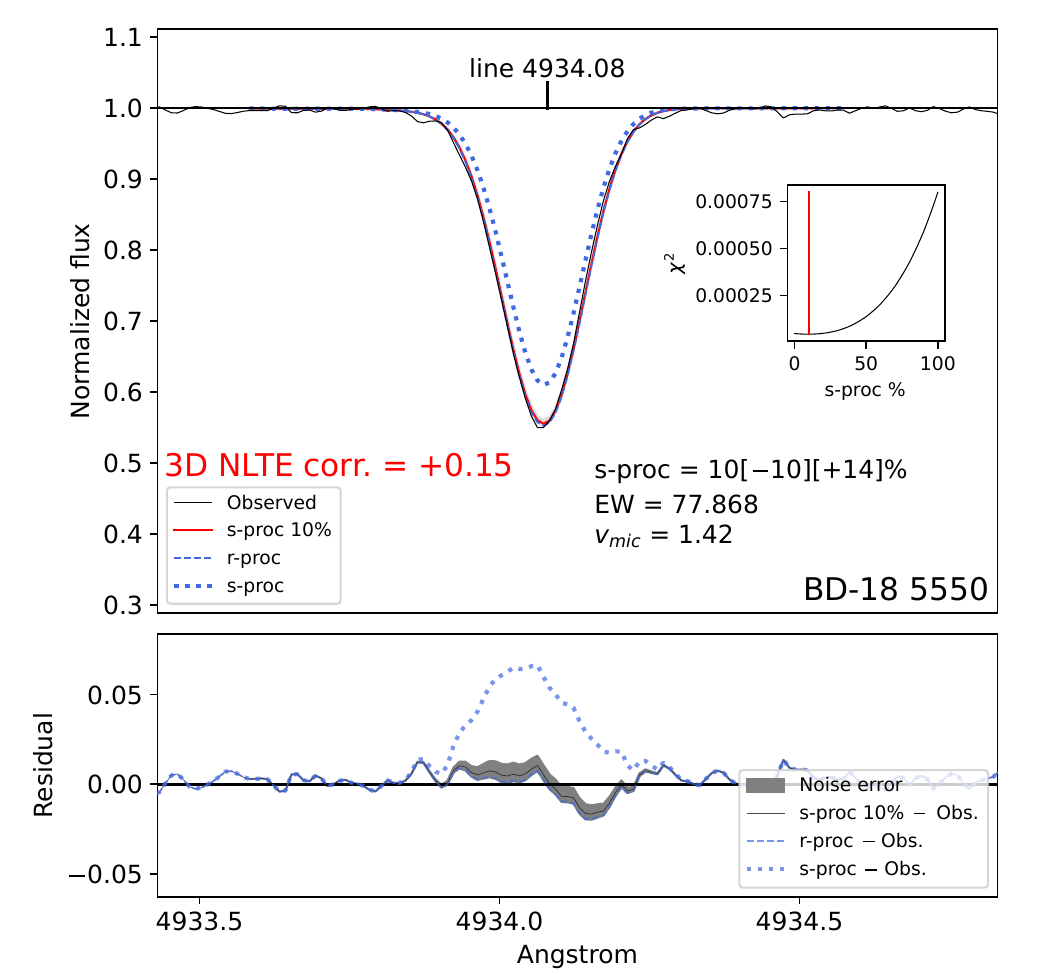}
    \includegraphics[width=0.4\linewidth]{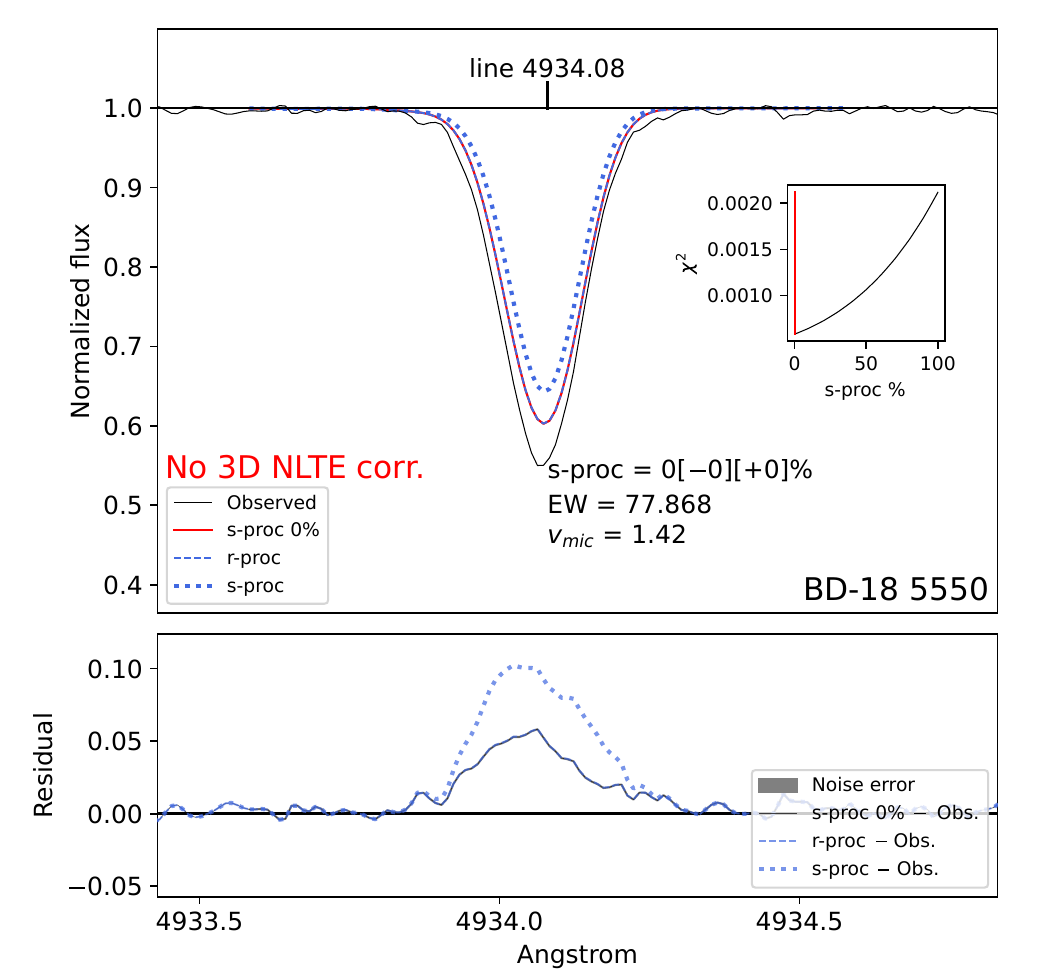}
    \caption{\tiny Similar to left panels in Fig.~\ref{fig:giant_fits}, but for BD$+18\,5550$.
    The fits in the left and right panels are done considering and neglecting 3D~NLTE corrections, respectively (see main text).
    }
    \label{fig:bd18}
\end{figure*}

BD$-18\,5550$ exhibits extremely low neutron-capture element abundances (Ba, La, Nd, Eu, Dy, Ho, Er, and Yb), with an overall abundance pattern consistent with an r-process origin according to \citep{roederer2017ApJ...835...23R}. 
They derived a relatively high [Ba/Eu] $= -0.41$~dex, suggestive of a weak r-process contribution \citep[e.g.][]{truran2002PASP..114.1293T}. 
Applying a 3D~NLTE correction of +0.05~dex to our Ba abundance yields [Ba/Eu] $= -0.84^{+0.14}_{-0.25}$~dex (from values in Table~\ref{tab:barium}), in close agreement with the main solar r-process component \citep[$-0.8$~dex,][]{Simmerer2004ApJ...617.1091S} and  consistent with the global abundance pattern discussed by \cite{roederer2017ApJ...835...23R}.
Figure~\ref{fig:bd18} shows the fit to the resonance line, for which an almost pure r-process isotopic mixture is diagnosed. 
The synthesis was computed adopting A(Ba) $= -1.44 + 0.05 + 0.10$~dex, where the latter two terms correspond to the 3D~NLTE corrections applied to the subordinate and resonance lines, respectively. 
Without these corrections, the observed line profile remains deeper than predicted by either the r- or s-process synthetic profiles, as exhibited in the right panel of the figure.

\section{[Ba/Ce], [Ba/La], [Ba/Sr] abundance ratios}
\label{sec:other_el}

\begin{figure}
    \centering
    \includegraphics[width=0.8\linewidth]{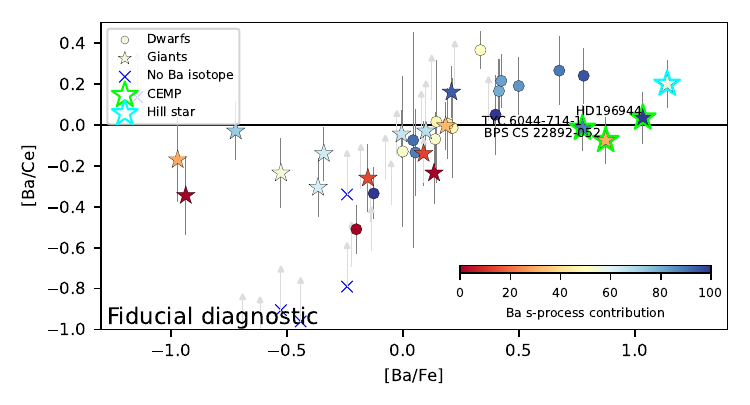}\\
    \includegraphics[width=0.8\linewidth]{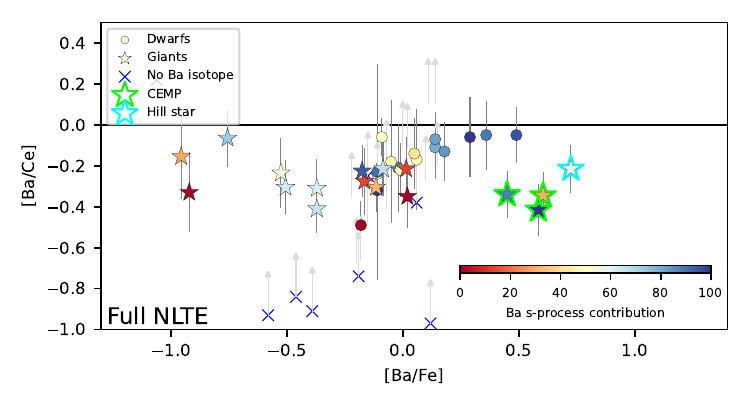}
    \caption{\tiny [Ba/Ce] versus [Ba/Fe] diagram. The elements of the plots are similar to those in Fig.~\ref{fig:BaEu_BaFe}. Gray arrows indicate stars with upper limits of Ce. Stars without isotopic ratio determination are indicated by blue crosses.
    }
    \label{fig:BaCe}
\end{figure}

\begin{figure}
    \centering
    \includegraphics[width=0.8\linewidth]{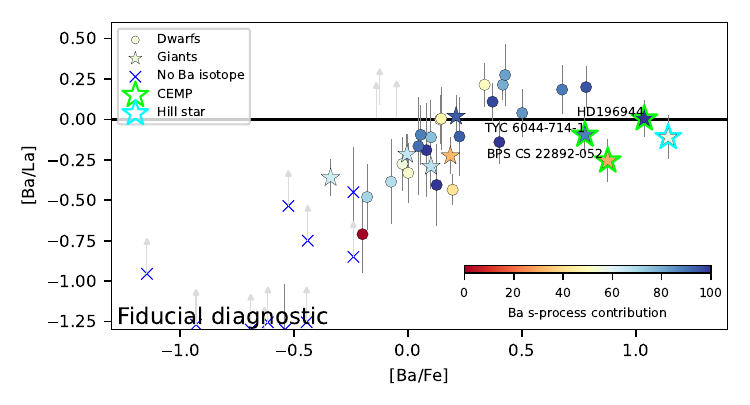}\\
    \includegraphics[width=0.8\linewidth]{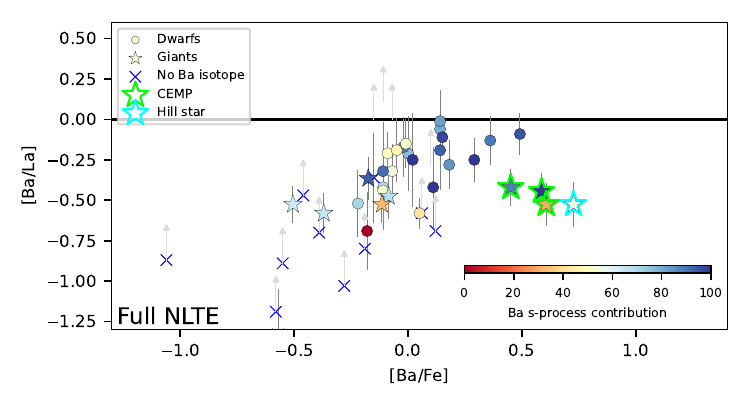}
    \caption{\tiny [Ba/La] versus [Ba/Fe] diagram. The elements of the plots are similar to those in Fig.~\ref{fig:BaEu_BaFe}. Gray arrows indicate stars with upper limits of La. Stars without isotopic ratio determination are indicated by blue crosses.
    }
    \label{fig:BaLa}
\end{figure}

\begin{figure}
    \centering
    \includegraphics[width=0.8\linewidth]{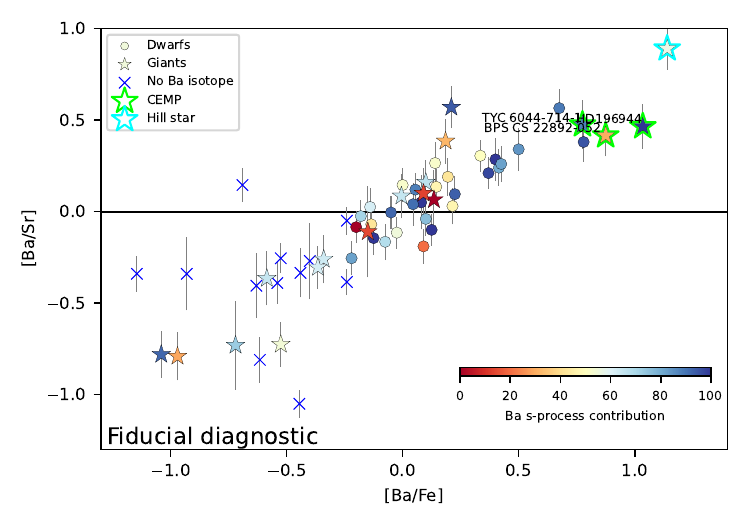}\\
    \includegraphics[width=0.8\linewidth]{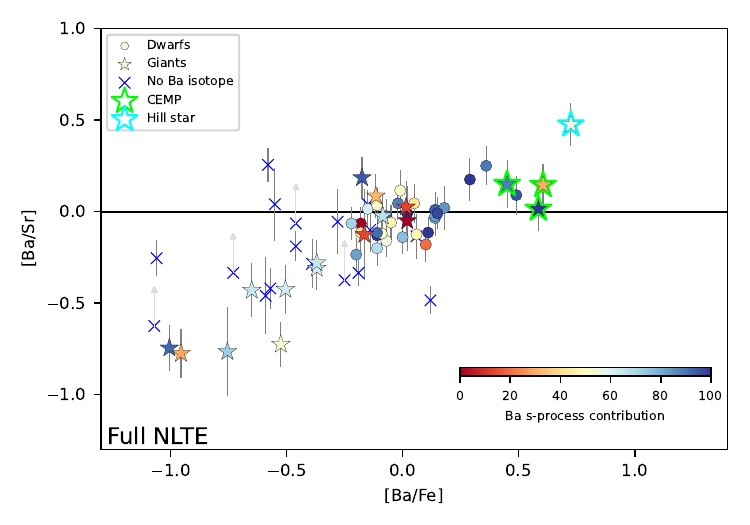}
    \caption{\tiny [Ba/Sr] versus [Ba/Fe] diagram. The elements of the plots are similar to those in Fig.~\ref{fig:BaEu_BaFe}. Gray arrows indicate stars with upper limits of Sr. Stars without isotopic ratio determination are indicated by blue crosses.
    }
    \label{fig:BaSr}
\end{figure}

Figures~\ref{fig:BaCe}, \ref{fig:BaLa}, and \ref{fig:BaSr} show the distributions of [Ba/Ce], [Ba/La], and [Ba/Sr] as functions of [Ba/Fe]. Our fiducial diagnostic indicates that these ratios vary from negative to positive values as [Ba/Fe] increases, as expected. In contrast, the full 1D~NLTE diagnostic yields exclusively negative [Ba/Ce] and [Ba/La] ratios and systematically low [Ba/Sr], regardless of the inferred s-process contribution to the Ba abundance. We interpret this behaviour as evidence that 1D~NLTE underestimates the Ba abundance.

In the fiducial diagnostic, [Ba/Ce], [Ba/La], and [Ba/Sr] approach zero at [Ba/Fe] $\sim 0.2$~dex. Most stars with positive abundance ratios are s-process dominated, although stars with mixed r- and s-process contributions may also exhibit positive ratios. The vertical dispersion is similar for [Ba/Ce] and [Ba/La], with an amplitude of about 0.7~dex. The observed trends likely reflect Galactic chemical evolution, as stars with higher [Ba/Fe] also tend to have higher [Fe/H] and younger ages.

For peculiar stars, where heavy elements are typically produced in one or a few enrichment events, the sign of [Ba/Ce] and [Ba/La] depends on the nucleosynthetic origin and properties of the event. For example, the s-process can produce either positive or negative [Ba/Ce] and [Ba/La] ratios depending on the initial mass of the AGB progenitor \citep{sbordone2020A&A...641A.135S}. In any case, although Ce and La are neighbouring elements to Ba, they cannot constrain the Ba abundance more precisely than the dispersion observed in Figs.~\ref{fig:BaCe} and \ref{fig:BaLa}, namely $\pm 0.25$~dex. As shown in Table~6 of \cite{giribaldi2025A&A...702A..65G}, such an uncertainty corresponds to approximately $\pm 70\%$ in the inferred s-process contribution when used to derive isotopic ratios from the 4934~\AA\ resonance line.

\begin{table}
\caption{Atomic parameters of fitted lines}
\label{tab:atomic_parameters}
\centering
\tiny 
\begin{threeparttable}
\begin{tabular}{lcccc}
\hline\hline
Element & centre (\AA) & $\chi_{\mathrm{ex}}$ & log~$gf$ & type\\
\hline
\ion{Eu}{II} & 4129.720 & 0.000 & 0.220 & d/g\\
\ion{Eu}{II} & 4205.03 & 0.000 & 0.210 & d/g\\
\ion{Ce}{II} & 3999.237 & 0.295 & 0.060 & g \\
\ion{Ce}{II} &  4073.474 & 0.478 & 0.210 & g\\
\ion{Ce}{II} &  4075.700 & 0.701 & 0.230 & g\\
\ion{Ce}{II} &  4075.847 & 0.609 & 0.160 & g\\
\ion{Ce}{II} & 4127.364 & 0.684 & 0.310 & g \\
\ion{Ce}{II} & 4137.645 & 0.516 & 0.400 & g\\
\ion{Ce}{II} & 4222.597 & 0.122 & $-0.150$ & d/g\\
\ion{Ce}{II} & 4562.359 & 0.478 & 0.210 & g\\
\ion{Ce}{II} & 5274.229 & 1.044 & 0.130 & g\\
\ion{Ce}{II} & 6043.373 & 1.206 & $-0.480$ & g \\
\ion{La}{ii} & 5114.560 & 0.235 & $-1.030$ & d/g\\
\ion{La}{ii} & 5122.984 & 0.321 &  $-1.111$ & d/g\\
\ion{La}{ii} & 5259.380 & 0.173 & $-1.950$ & d/g\\
\ion{La}{ii} & 5290.820 & 0.000 & $-1.650$ & d/g\\
\ion{La}{ii} & 5303.514 & 0.321 & $-1.731$ & d/g\\
\ion{La}{ii} & 5936.220 & 0.173 &  $-2.809$ & d/g\\
\ion{La}{ii} & 6390.486 & 0.321 & $-1.903$ & d/g\\
\ion{Sr}{ii} & 3464.453 & 3.040 & 0.489 & g\\
\ion{Sr}{ii} & 3474.889 & 3.040 & $-0.467$ & g\\
\ion{Sr}{ii} & 4077.709 & 0.000 & 0.143 & d \\
\ion{Sr}{ii} & 4161.792 & 2.940 & $-0.327$ & d/g\\
\ion{Sr}{ii} & 4215.519 & 0.000 & $-0.173$ & d\\
\hline
\end{tabular}
\begin{tablenotes}
\item{} \textbf{Notes.} {The rightmost column indicates the type of the star for which the line was used.
} 
\end{tablenotes}
\end{threeparttable}
\end{table}

\begin{figure}
    \centering
    \includegraphics[width=1\linewidth]{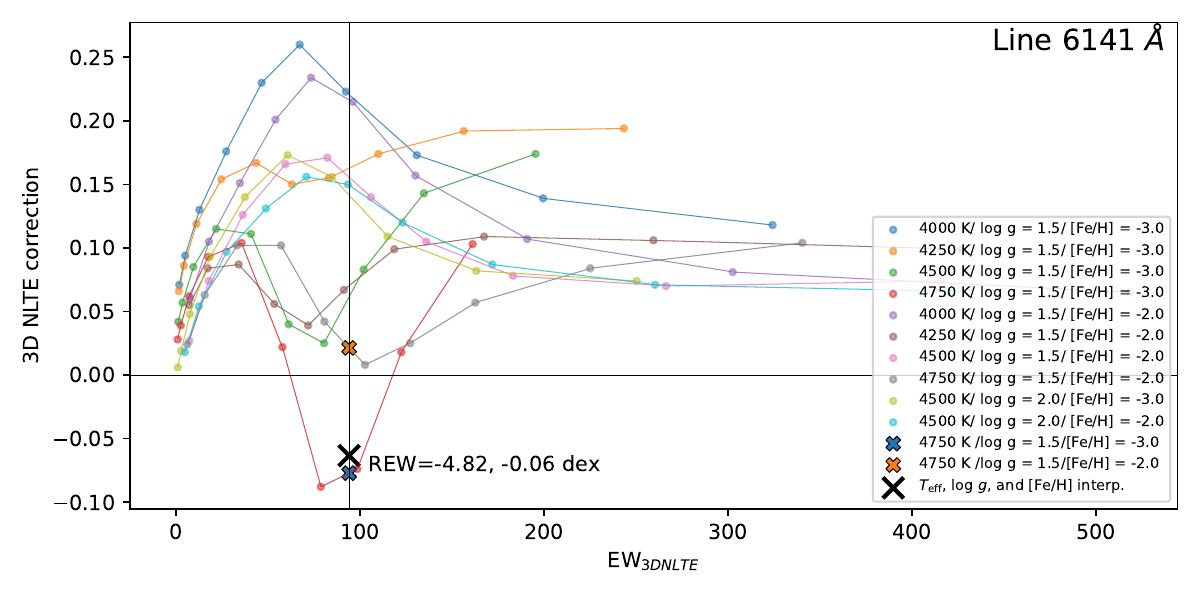}
    \caption{\tiny Interpolation of the 3D~NLTE corrections for the $\lambda$6141~\AA\ line in the star BPS~CS~22892-052. The circles correspond to the tabulated values from the original grids, with their associated stellar parameters indicated in the legend. Solid lines in matching colours show the linear interpolations between these grid points. Cross symbols mark the values obtained by interpolation in \teff\ and \logg, while the black cross indicates the result derived from interpolation in \teff, \logg, and [Fe/H]. The figure also labels the line’s REW and the resulting 3D~NLTE correction.
    }
    \label{fig:3Dcors}
\end{figure}

\section{Additional stochastic CEM predictions}

\subsection{Impact of AGB in the early Ba enrichment}

{In Section \ref{sss:model_res}, we use the [Ba/Fe] vs. [Fe/H] abundance diagram to get insights on the massive stars impact on the Ba s-process production, justifying this approach by the negligible enrichment by AGB in the early / metal-poor phase of chemical evolution.

\begin{figure}
    \centering
    \includegraphics[width=0.8\linewidth]{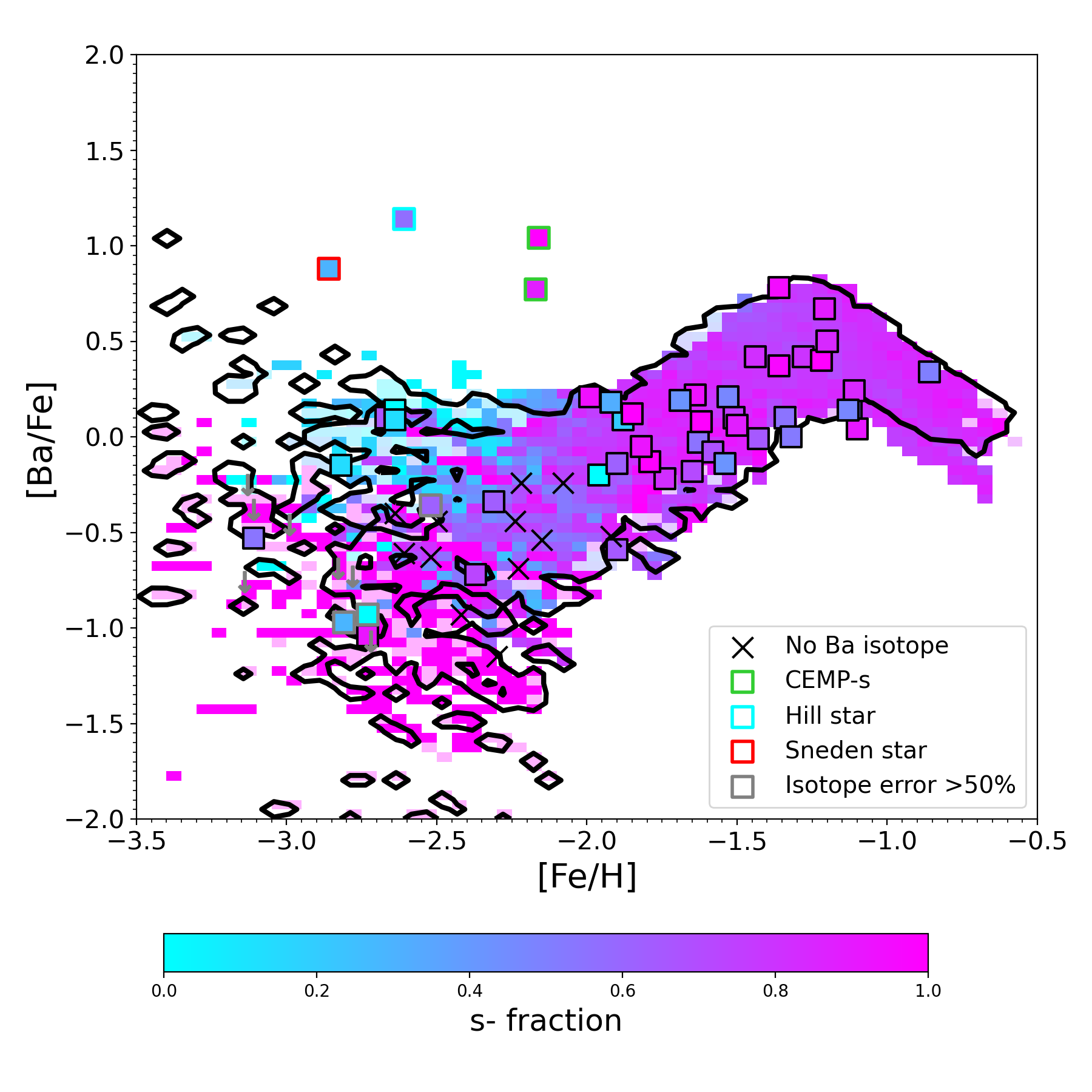}
    \caption{\tiny  [Ba/Fe] versus [Fe/H] diagram as predicted by a stochastic chemical evolution model switching-off the AGB contribution to s-process. Black contours represent the predictions by the fiducial model, while the colormap displays the fractional s- process contribution to Ba abundance as predicted by the model without AGB contribution Data legend is as in Fig. \ref{fig:BaFe_sfract}.
    }
    \label{fig:BaFe_AGB}
\end{figure}

For this reason, in Fig. \ref{fig:BaFe_AGB} we compare the results by our fiducial MW model with a run with analogue setup, except for the absence of AGB polluters for the s-process. 
The Figure well justifies the claims advocated in the main text. Indeed, even by switching-off AGBs, the predicted abundance patterns remain nearly identical, with a slight deviation only at [Fe/H]$\gtrsim$-1 dex, i.e. outside of the metallicity range displayed by the data sample. }

\subsection{Complementary abundance diagrams}

Below, we present stochastic CEM results for abundance ratios linked to those presented in Section \ref{sss:model_res} figures. 
In particular, we display the results for [Eu/Fe] versus [Fe/H] and [Ba/Eu] vs. [Fe/H]. 
The reproduction of the patterns observed for these abundance ratios by models is very important, as it allows to better probe model consistency across elements produced by different nucleosynthetic origin (e.g. pure r-process for Eu) and therefore check whether the patterns shown in Section \ref{sss:model_res} are not the result of spurious predicted patterns.

\begin{figure}
    \centering
    \includegraphics[width=0.8\linewidth]{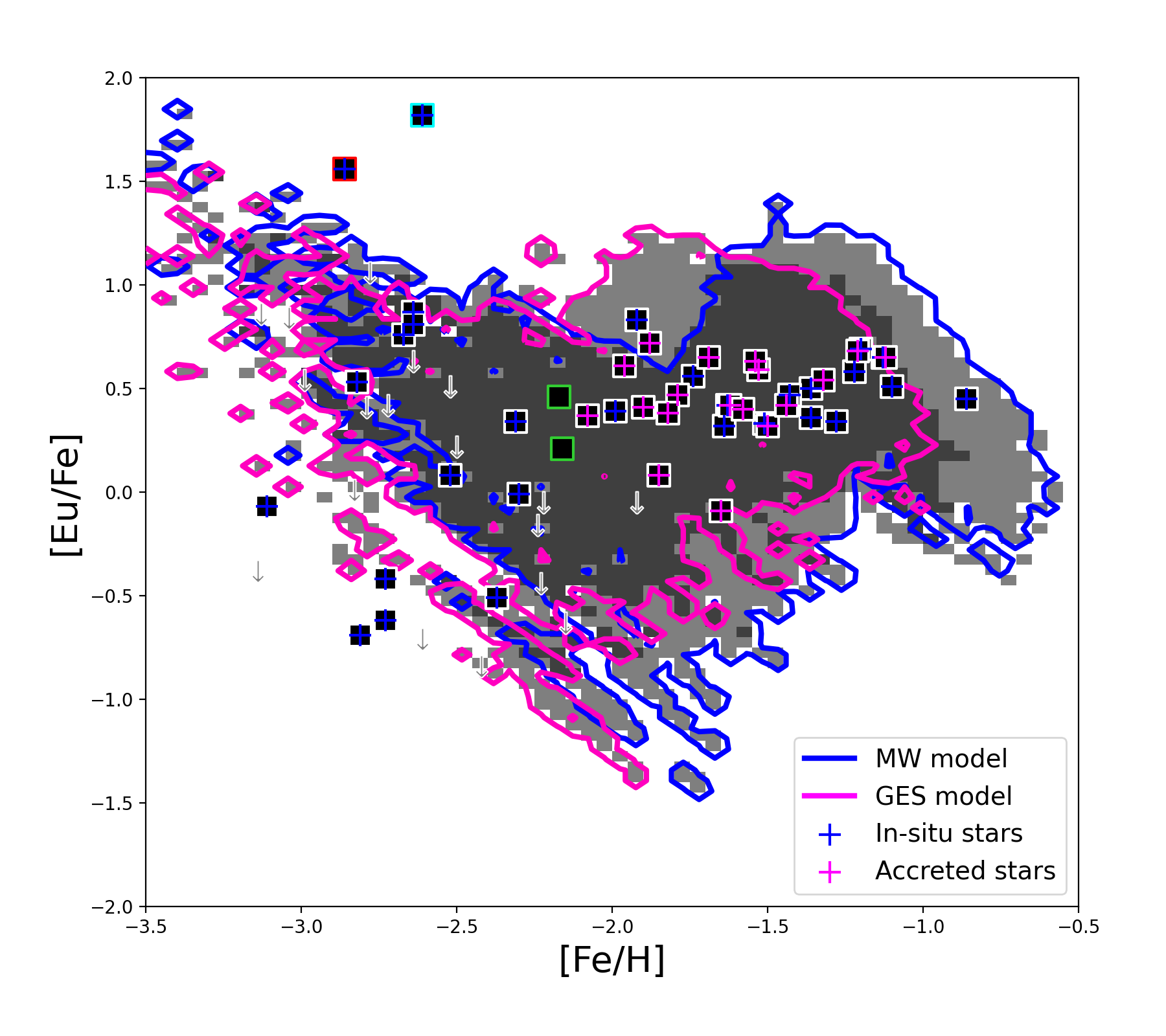}
    \caption{\tiny [Eu/Fe] versus [Fe/H] diagram as predicted by our fiducial stochastic chemical evolution model framework. Blue contours identify the MW predictions, while magenta contours the GES model. Data for MW and GES member stars are identified by blue and magenta crosses, respectively. CEMP-s stars, Hill star and Sneden star are labelled as in Fig. \ref{fig:BaFe_sfract}.
    }
    \label{fig:EuFe}
\end{figure}

In Figure \ref{fig:EuFe}, we show the predictions for [Eu/Fe] vs. [Fe/H] for the early MW and GES models as compared with the data. Both the models well reproduce the general trend and scatter observed for normal stars across different metallicities. 
Only few Eu-poor ([Eu/Fe]~$<0$ dex) stars with metallicites [Fe/H]~$\lesssim-2.5$ dex struggle in being reproduced by the models, in particular that of the MW. This happens despite introducing a scatter in r-process production (see Eq. \eqref{eq:rproc_scatter}), which prevents producing exclusively supersolar [Eu/Fe] at [Fe/H]~$\lesssim-2.5$~dex. 
Several explanations can be advocated to explain such (limited) mismatch, such as partial retention by the galaxy of the Eu produced in energetic events as MRD-SNe or NSM \citep[see][]{Bonetti19,Cavallo23} or large variability in Fe yields by the first, low-metallicity SNe (see \citealt{Tominaga14} and references therein).
In any case, a careful inclusion of all these effects will add further layers of complexity to the adopted, well-test model setup, without a significant benefit on the data-model comparison.
Indeed, the scope of this paper is to probe the production of Ba r- and s-process sources in a large range of metallicities, and those metal-poor, Eu-poor stars constitute a small minority across the entire sample.

\begin{figure}
    \centering
    \includegraphics[width=0.8\linewidth]{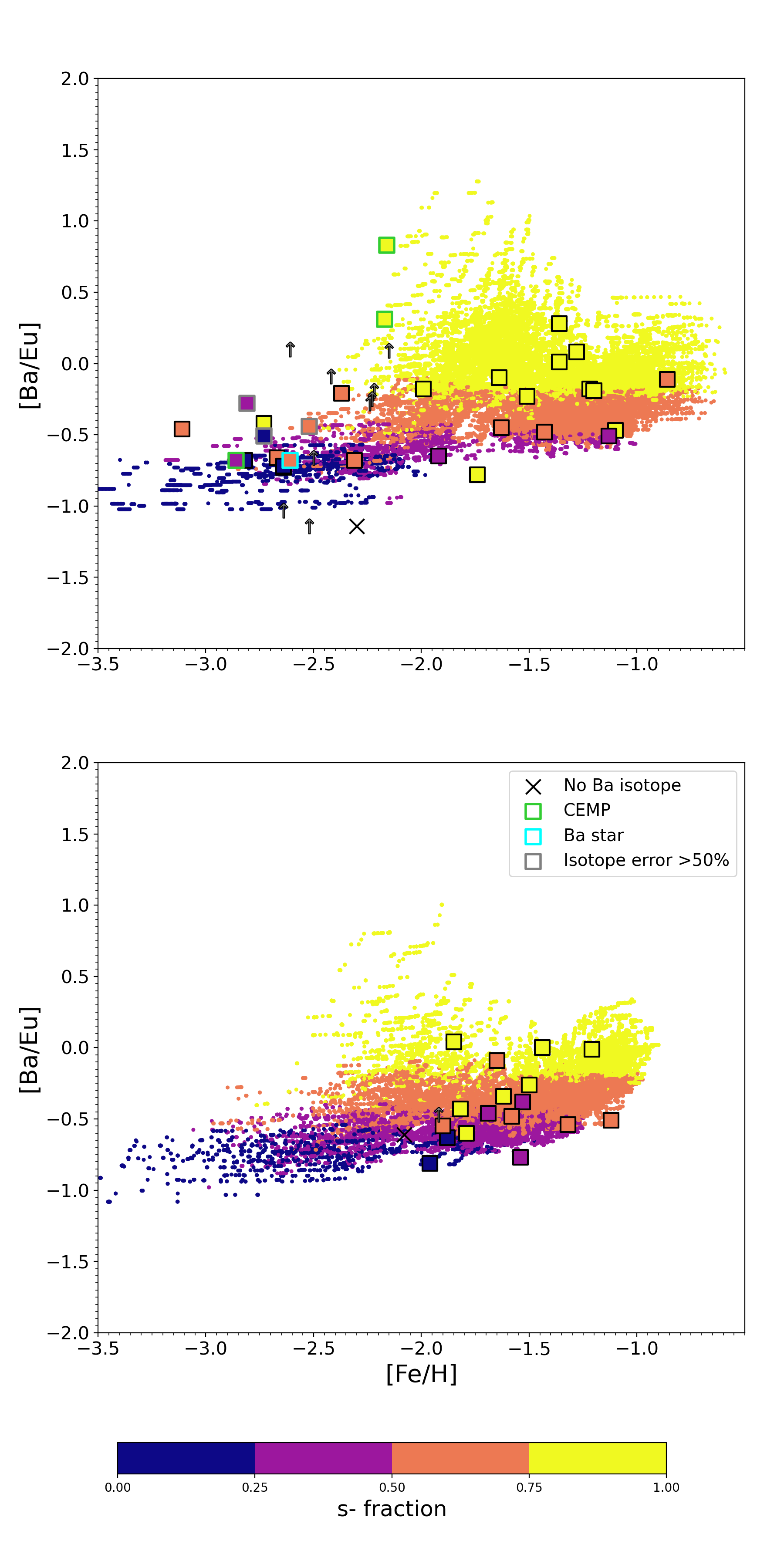}
    \caption{\tiny [Ba/Eu] versus [Fe/H] diagram as predicted by our fiducial stochastic chemical evolution model framework. Top panel shows predictions by the MW model, bottom panel for the GES model. For both panels, the colormap displays the fractional s-process contribution to Ba abundance. Data legend is as in Fig.~\ref{fig:BaEu_BaFe_sfract}.
    }
    \label{fig:BaEu_FeH_sfract}
\end{figure}

Figure \ref{fig:BaEu_FeH_sfract} shows the evolution of [Ba/Eu] as function of [Fe/H] for both the MW model (top panel) and the GES model (bottom panel). 
It is clear from this Figure that the presence of such small subsample of metal-poor, Eu-poor stars do not prevent to capture the trend of [Ba/Eu] with the fraction of Ba produced by s-process, both generally increasing with metallicity. 
In addition, it is worth noting that models well reproduce the incremental scatter of both [Ba/Eu] and s-fraction with [Fe/H], further confirming the consistency of the model in reproducing the different features of different observables.

\end{appendix}

\end{document}